\documentclass[american,aps,pre,twocolumn]{revtex4-1}
\usepackage[T1]{fontenc}
\usepackage[latin9]{inputenc}
\usepackage{color}
\usepackage{array}
\usepackage{verbatim}
\usepackage{textcomp}
\usepackage{amsmath}
\usepackage{amssymb}
\usepackage{stmaryrd}
\usepackage{graphicx}
\usepackage{wasysym}
\usepackage{esint}

\makeatletter

\providecommand{\tabularnewline}{\\}

\makeatother

\usepackage{babel}
\begin{document}

\title{Virial series for inhomogeneous fluids applied to the Lennard-Jones
wall-fluid surface tension at planar and curved walls}

\author{$^{*\dagger}$Ignacio Urrutia}

\email{iurrutia@cnea.gov.ar}

\selectlanguage{american}%

\author{$^{*\dagger}$Iván E. Paganini}

\affiliation{$^{*}$Departamento de Física de la Materia Condensada, Centro Atómico
Constituyentes, CNEA, Av.Gral.~Paz 1499, 1650 Pcia.~de Buenos Aires,
Argentina}

\affiliation{$^{\dagger}$CONICET, Avenida Rivadavia 1917, C1033AAJ Buenos Aires,
Argentina}
\begin{abstract}
We formulate a straightforward scheme of statistical mechanics for
inhomogeneous systems that includes the virial series in powers of
the activity for the grand free energy and density distributions.
There, cluster integrals formulated for inhomogeneous systems play
a main role. We center on second order terms that were analyzed in
the case of hard-wall confinement, focusing in planar, spherical and
cylindrical walls. Further analysis was devoted to the Lennard-Jones
system and its generalization the 2k-k potential. For this interaction
potentials the second cluster integral was evaluated analytically.
We obtained the fluid-substrate surface tension at second order for
the planar, spherical and cylindrical confinement. Spherical and cylindrical
cases were analyzed using a series expansion in the radius including
higher order terms. We detected a $\ln R^{-1}/R^{2}$ dependence of
the surface tension for the standard Lennard-Jones system confined
by spherical and cylindrical walls, no matter if particles are inside
or outside of the hard-walls. The analysis was extended to bending
and Gaussian curvatures, where exact expressions were also obtained.
\end{abstract}
\maketitle

\section{Introduction\label{sec:Intro}}

Equation of state (EOS) of a bulk fluid system contains the information
about its thermodynamic behavior. For known potentials, virial expansion
is a common method to calculate the EOS. This approach is usually
limited to a certain low density region such as gas phase and must
avoid transitions where the method is expected to break down. Additional
problems comes from series convergence itself. Virial series are
central for statistical mechanics (SM) theoretical development, hence
they are a recurring topic even after 150 years.\cite{vanderWaals_1873,vanLaar_1899,Ushcats_2013,Hellmann_2011,Shaul_2012}
There are several procedures that enable to obtain the EOS and other
properties of the fluids by extrapolation of the first known terms
of the virial series. Also, the first virial coefficients are used
in the development of liquid theories, like density functionals or
integral-differential equations.\cite{Gazzillo_2013,BeltranHeredia_2014,Korden_2012}

Virial series are not only a major tool for simple and molecular fluids,
but for colloidal systems. This systems are a mixture of two type
of particles with a characteristic big size difference.\cite{Dudowicz_2015,Koga_2015,LopezdeHaro_2015}
Nowadays, virial coefficients and cluster integrals are still under
study, even at lower orders. There are several recent works about
second order coefficients for diverse systems including inert gases,
alkanes, methane in water solution, and polymer solutions.\cite{Hutem_2012,Schultz_2010,Ashbaugh_2015,Dudowicz_2015,Koga_2015}
Distinct interaction models have been recently analyzed in this context:
hard spheres with dipolar momentum,\cite{Virga_2013,HendersonD_2011,Philipse_2010}
exponential potential,\cite{Mamedov_2015} and the Asakura-Oosawa
model for colloids.\cite{Santos_2015,LopezdeHaro_2015,Koga_2015,Dudowicz_2015}

One of the most studied interaction models for simple fluids is the
LJ. Modern studies based on molecular dynamics simulation shed light
on its basic properties as viscosity, thermal conductivity, cavitation
and melting coexistence.\cite{Baidakov_2012,Baidakov_2014c,Baidakov_2014,Baidakov_2014b,Heyes_2015}
Other works have focused on the curvature dependence of the surface
tension.\cite{Blokhuis_2013c,Blokhuis_2007,vanGiessen_2002} The virial
coefficients of the LJ fluid have been calculated numerically up to
sixteenth order \cite{Gibbons_1971,Caligaris_1971,Schultz_2009_b,Feng_2015}
and similar studies were done in LJ fluid mixtures up to sixth order.\cite{Schultz_2009}
Second order coefficient is particularly relevant in this work. It
was evaluated exactly for the first time in 2001 by Vargas et al.\cite{Vargas_2001}
and re-evaluated later.\cite{Eu_2009,Mamedov_2014} Generalizations
to the so-called 2k-k LJ system\cite{Glasser_2002} and extensions
to non-conformal LJ model, were also done.\cite{GonzalezCalderon_2015}
We can mention that both, simple and colloidal fluids are continuously
studied because some of their properties are yet not completely understood,
being the 2k-k LJ interaction one of the models that enable to analyze
them in a unified framework. 

All the mentioned works about virial series refer to homogeneous fluids.
In fact, most of the theoretical development about virial series is
based on the original formulation and thus only apply to homogeneous
systems.\cite{Mayer1940,Hill1956,McQuarrie2000,Hansen2006} Later
generalizations adapted virial series expansions to inhomogeneous
fluids and include external potentials. The seminal work on inhomogeneous
systems was done by Bellemans in the sixties.\cite{Bellemans_1962,Bellemans_1962_b,Bellemans_1963}
Further developments were done by Sokolowski and Stecki,\cite{Sokolowski_1977,Sokolowski_1979,Stecki_1980}
and by Rowlinson.\cite{Rowlinson_1986,Rowlinson_1985}

In this work we briefly introduce in a simple manner the statistical
mechanics approach to inhomogeneous systems in grand canonical ensemble
and its virial series. Our presentation focuses on a system of particles
confined by the action of a general external potential following Rowlinson's
approach. We discuss virial series at the level of power series in
the activity, where cluster integrands and integrals play a central
role. To make simpler both notation and explanations the treatment
is based on a one component system, and to some extent, to particles
with pair-additive interaction. Despite this, extensions to mixtures
including polyatomic molecules with internal degrees of freedom, and
generalizations beyond the two-body potential that enable inclusion
of multibody interactions are discussed. Alongside, our treatment
of free energy, density distributions and other properties avoids
the necessity of a volume definition. All the questions related to
establish the volume and a reference bulk homogeneous system are also
left to a separate analysis.

As an application we analyze the terms of second order for spherically
symmetric pair interaction potentials. We solved for the first time
the second order cluster integral for LJ and 2$k$-$k$ LJ fluids
under inhomogeneous conditions. We evaluate analytically the cluster
integral in non-trivial confinements: those produced by planar, spherical
and cylindrical hard walls. Our expression is exact for the planar
case. For curved walls we obtain several terms of the asymptotic expansion
for large radii. To highlight the difficulty of the actual problem,
we mention that up to date the only cluster integral analytically
solved is that of second order and for the bulk case. This term corresponds
to the pressure second virial coefficient.

Using the expression for the second cluster integral we study the
properties of the LJ gas in contact with a curved hard wall, focusing
on its surface tension. The question of how the properties of an inhomogeneous
fluid depend on the curvature of its interface is a long standing
problem in statistical mechanics.\cite{Blokhuis_2013c} It has been
thoroughly studied even for the interface induced by a curved substrate.\cite{Evans_2003,Evans_2004,Stewart_2005,Stewart_2005_b,Blokhuis_2007,Reindl_2015}
We present rigorous results based on virial series about the curvature
dependence of the properties of the LJ gas in contact with a curved
wall which are exact up to power two in density. 

The rest of this work is organized as follows: In Sec. \ref{sec:Theory}
the SM of open systems with fixed chemical potential, temperature
and external potential are revisited. Cluster integrals are thus shown
in their inhomogeneous nature. Second order cluster terms are analyzed
in Sec.\ref{sec:SndOrderTerms}. There, the case of spherically interacting
particles lying in a spatial region of arbitrary shape where they
freely move is analyzed. We emphasize on three types of simple geometry
confinements: planar, spherical and cylindrical walls. An application
for the Lennard-Jones and the generalized 2$k$-$k$ Lennard-Jones
systems is given in Sec.\ref{sec:Apps}, where analytic expressions
for the second cluster integral are derived. Sec. \ref{sec:Results}
is devoted to analyze the inhomogeneous low density gas, its wall-fluid
surface tension, Tolman length and rigidity coefficients of bending
and Gaussian curvatures. Finally, in Sec. \ref{sec:Summary} we give
our conclusions and final remarks.

\section{Theory\label{sec:Theory}}

We consider an inhomogeneous system at a given temperature $T$, chemical
potential $\mu$ (the number of particles may fluctuate) and external
potential. The total potential energy also includes the contribution
of mutual interaction between particles $\phi_{(n)}$.\textcolor{cyan}{{}
}Thus, the grand canonical ensemble partition function (GCE) is 
\begin{equation}
\Xi=1+\sum_{n=1}\lambda^{n}Q_{n}\:,\label{eq:gcO}
\end{equation}
where $\lambda=\exp(\beta\mu)$ and $\beta=1/kT$ is the inverse
temperature ($k$ is the Boltzmann's constant). In Eq. (\ref{eq:gcO})
$Q_{n}$ is the canonical ensemble partition function
\begin{eqnarray}
Q_{n} & = & \Lambda^{dn}Z_{n}/n!\:,\label{eq:Qn-1}\\
Z_{n} & = & \int g_{n}\left(\mathbf{x}\right)\exp\left(-\beta\phi_{(n)}\right)d\mathbf{x}\:,\label{eq:Zn-1}
\end{eqnarray}
where $\Lambda$ is the de Broglie thermal wavelength, $d$ is dimensionality
and $Z_{n}$ is the configuration integral. $g_{n}\left(\mathbf{x}\right)=\prod_{i=1}^{n}g\left(\mathbf{x}_{i}\right)$,
$g\left(\mathbf{x}_{i}\right)=\exp\left(-\beta\psi_{i}\right)$ and
$\psi_{i}$ is the external potential over the particle $i$.

In Eq. (\ref{eq:gcO}) the sum index may end both at a given $m$
representing the maximum number of particles in the open system or
at infinity. Fixing the value of $m$ allows the study of small systems.\cite{Rowlinson_1986}
The main link between GCE and thermodynamics is still
\begin{equation}
\beta\Omega=-\ln\Xi\:.\label{eq:OmgLog}
\end{equation}
Some thermodynamic magnitudes could be directly derived from $\Omega$
as $\left\langle n\right\rangle =-\beta\lambda\partial\Omega/\partial\lambda$.
Yet, other thermodynamic magnitudes could be derived from $\Omega$
once volume and area measures of the system are introduced.

In the GCE several magnitudes can be expressed as power series in
the activity $z=\lambda/\Lambda^{3}$ (virial series in $z$), with
cluster integrals $\tau_{k}$ and cluster residual part as coefficients.
The most frequent in the literature are 
\begin{eqnarray}
\beta\Omega & = & -\sum_{k=1}^{\infty}\frac{z^{k}}{k!}\tau_{k}\:,\label{eq:Omgz}\\
\left\langle n\right\rangle  & = & \sum_{k=1}^{\infty}\frac{kz^{k}}{k!}\tau_{k}\:,\label{eq:meanN}\\
\delta n & = & \left\langle n^{2}\right\rangle -\left\langle n\right\rangle ^{2}=\sum_{k=1}^{\infty}\frac{k^{2}z^{k}}{k!}\tau_{k}\:.
\end{eqnarray}
Here $\left\langle n\right\rangle $ is the mean number of particles
in the system and $\delta n$ measures the fluctuation of $\left\langle n\right\rangle $.
Cluster integrals have played an important role in the development
of virial expansion for homogeneous systems. For inhomogeneous fluids
it is convenient to define the $n$-particles cluster integral $\tau_{n}$
as
\[
\tau_{n}=n!\int g_{n}(\mathbf{x})\,b_{n}\left(\mathbf{x}_{1},\ldots,\mathbf{x}_{n}\right)d\mathbf{x}\:,
\]
where $b_{n}\left(\mathbf{x}_{1},\ldots,\mathbf{x}_{n}\right)$ is
the Mayer's cluster integrand of order $n$. For simplicity, from
here on we assume a pair potential interaction i.e. $\phi_{(n)}=\sum_{i,j}\phi_{ij}$,
$\phi_{ij}=\phi\left(\mathbf{\mathbf{x}}_{ij}\right)$, being $\mathbf{\mathbf{x}}_{ij}=\mathbf{\mathbf{x}}_{j}-\mathbf{\mathbf{x}}_{i}$
the vector between $i$ and $j$ particles. Thus, $b_{n}\left(\mathbf{x}_{1},\ldots,\mathbf{x}_{n}\right)$
is the sum of all the product of Mayer's function $f\bigl(\mathbf{\mathbf{x}}\bigr)=\exp\bigl[-\beta\phi\bigl(\mathbf{\mathbf{x}}\bigr)\bigr]-1$
that involves $n$ particles linked by $f$-bonds. In this sense $b_{n}\left(\mathbf{x}_{1},\ldots,\mathbf{x}_{n}\right)$
corresponds to clusters of $n$ particles.\cite{Yang_2013} Note that
$b_{1}=1$ and $\tau_{1}=Z_{1}$. For the density distributions the
same approach is also useful. The one body density distribution is\cite{Hill1956,Hansen2006}
\[
\rho\bigl(\mathbf{r}\bigr)=\bigl\langle\sum_{i=1}\delta\left(\mathbf{r}-\mathbf{x}_{i}\right)\bigr\rangle\:.
\]
It is convenient to define the residual or $n$-cluster part of $\rho\bigl(\mathbf{r}\bigr)$,
given by 
\[
\left\llbracket \rho^{(n)}\bigl(\mathbf{r}\bigr)\right\rrbracket =\frac{n}{Z_{n}}g_{1}\left(\mathbf{r}\right)\int g_{n-1}\left(\mathbf{x}\right)\,b_{n}\left(\mathbf{r},\mathbf{x}_{2},\ldots,\mathbf{x}_{n}\right)d\mathbf{x}\:,
\]
which plays the role that cluster integrals do in Eq. (\ref{eq:Omgz}).
The virial series for $\rho\bigl(\mathbf{r}\bigr)$\cite{McQuarrie2000}
is
\begin{eqnarray}
\rho(\mathbf{r}) & = & \sum_{k=1}^{\infty}\frac{z^{k}}{k!}Z_{k}\left\llbracket \rho^{(k)}(\mathbf{r})\right\rrbracket \:,\nonumber \\
 & = & \sum_{k=1}^{\infty}\frac{kz^{k}}{k!}\overline{\left\llbracket \rho^{(k)}(\mathbf{r})\right\rrbracket }\:.\label{eq:rho1}
\end{eqnarray}
Extension to other distribution functions are also direct. For example,
for the two body distribution function one has\cite{Hansen2006}
\[
\rho_{2}(\mathbf{r}_{1},\mathbf{r}_{2})=\bigl\langle\sum_{i=1}^{n}\sum_{j\neq i}^{n}\delta\left(\mathbf{r}_{1}-\mathbf{x}_{i}\right)\delta\left(\mathbf{r}_{2}-\mathbf{x}_{j}\right)\bigr\rangle\:,
\]
\begin{multline*}
\left\llbracket \rho_{2}^{(n)}(\mathbf{r}_{1},\mathbf{r}_{2})\right\rrbracket =\frac{n(n-1)}{Z_{n}}g_{1}(\mathbf{r}_{1})g_{1}(\mathbf{r}_{2})\exp\left[-\beta\phi(\mathbf{r}_{12})\right]\\
\times\int g_{n-2}(\mathbf{x})\,b_{n}\left(\mathbf{r}_{1},\mathbf{r}_{2},\mathbf{x}_{3},\ldots,\mathbf{x}_{n}\right)d\mathbf{x}\:,
\end{multline*}
\[
\rho_{2}(\mathbf{r}_{1},\mathbf{r}_{2})=\sum_{k=2}^{\infty}\frac{z^{k}}{k!}Z_{k}\left\llbracket \rho_{2}^{(k)}(\mathbf{r}_{1},\mathbf{r}_{2})\right\rrbracket \:.
\]
Treating particles 1 and 2 as linked, then $b_{n}\left(\mathbf{r}_{1},\mathbf{r}_{2},\mathbf{x}_{3},\ldots,\mathbf{x}_{n}\right)$
is the sum of all the cluster contributions of $n$ particles.\cite{McQuarrie2000}.

For homogeneous systems $g(\mathbf{x})=1$ and therefore $b_{n}(\mathbf{x})$
does not depend on the position, it reduces to the usual Mayer cluster
coefficient $b_{n}$. Thus performing an extra integration
\begin{equation}
\tau_{n}=n!\int_{\infty}b_{n}(\mathbf{r})d\mathbf{r}=n!Vb_{n}\:,\label{eq:Taunh}
\end{equation}
with $V$ the volume of the accessible region, i.e., the infinite
space or the cell when periodic boundary conditions are used.\cite{Hill1956}

We have restricted our cluster decomposition to the case of two body
interaction potentials. However, in Ref. \cite{Hellmann_2011} a systematic
analysis of many body terms was done for virial expansions in homogeneous
systems. It seems that with some minimal changes this approach is
applicable to the inhomogeneous case. On the other hand, second order
terms discussed in the next Sec. \ref{sec:SndOrderTerms} are not
modified when many-body interaction between particles are contemplated.

\section{Second order terms\label{sec:SndOrderTerms}}

\begin{figure}
\begin{centering}
\includegraphics[clip,width=0.26\columnwidth]{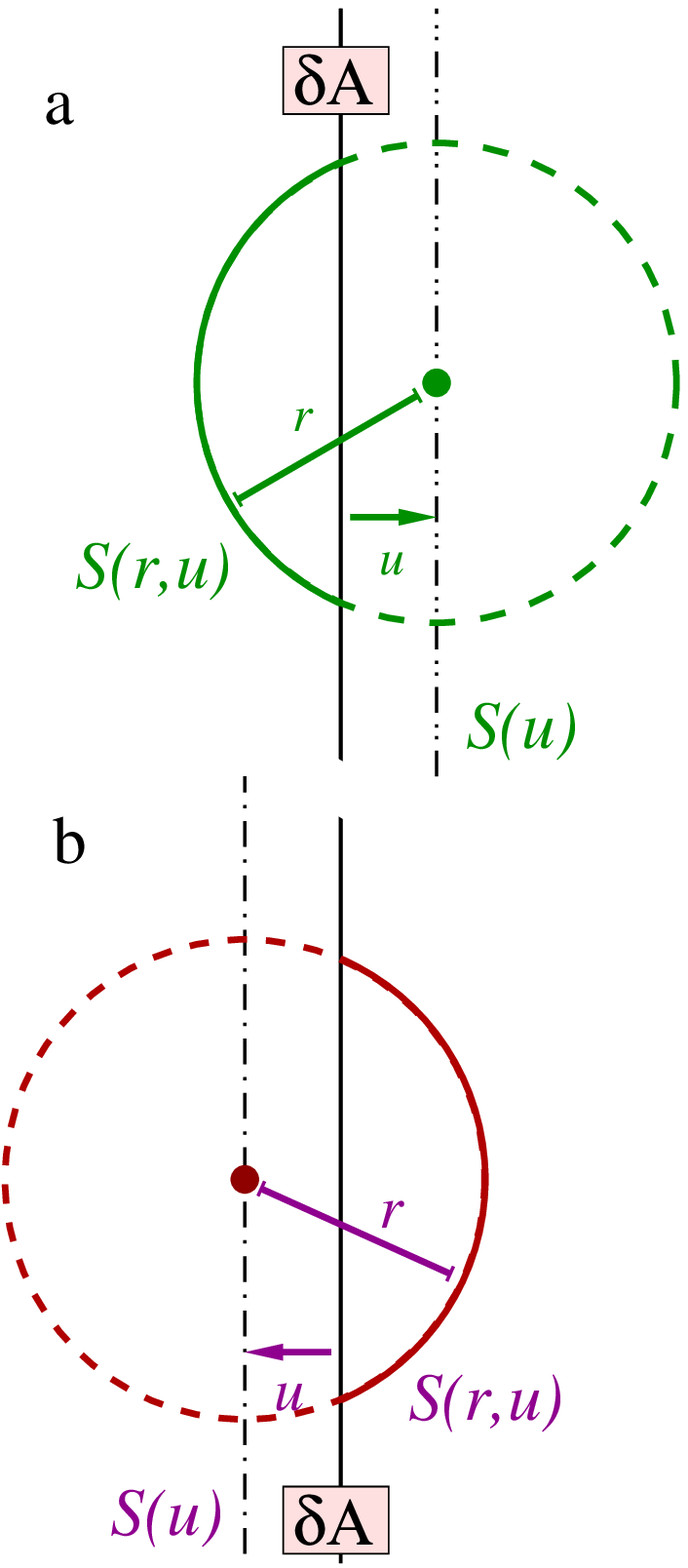} \includegraphics[bb=0bp 30bp 547bp 500bp,clip,width=0.72\columnwidth]{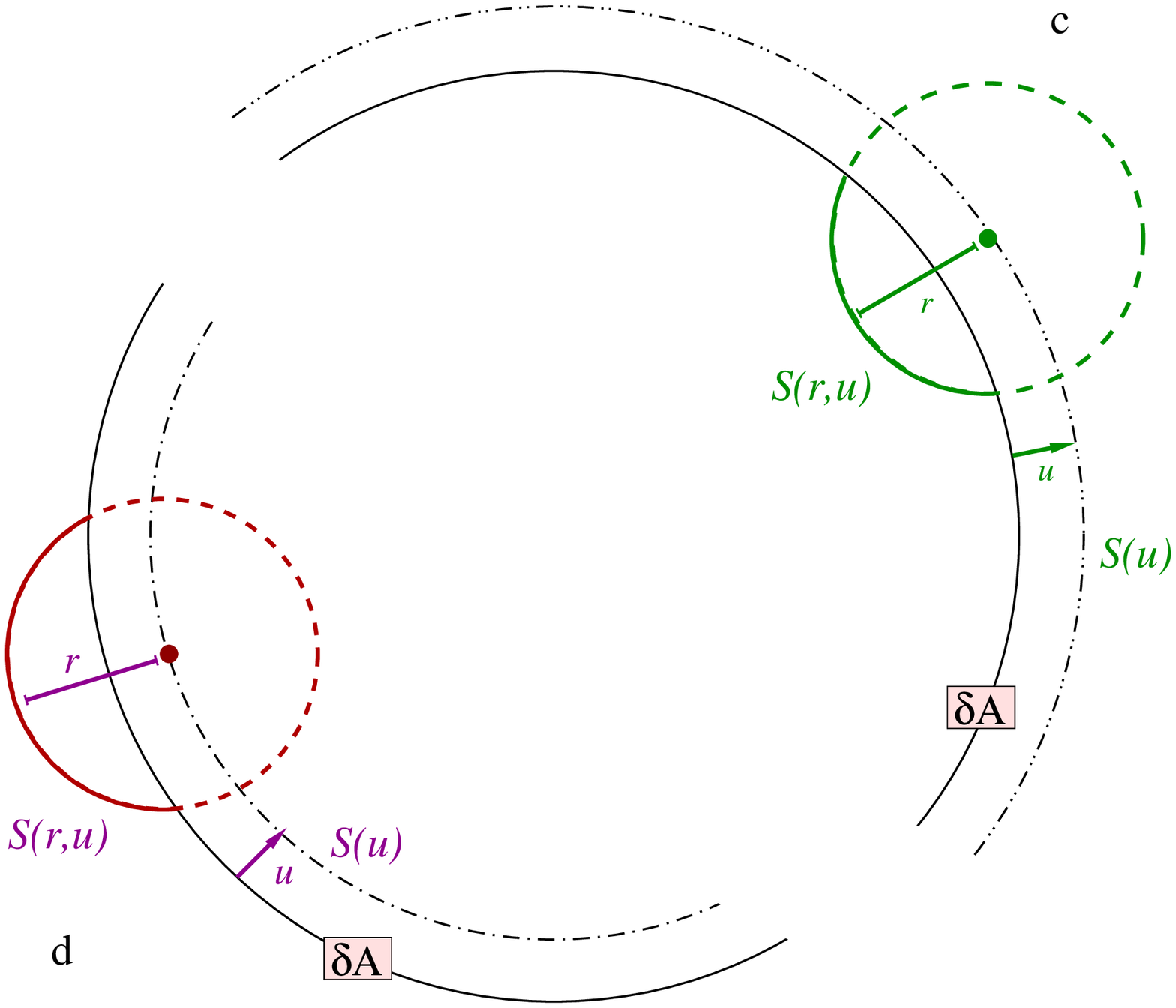}
\par\end{centering}

\caption{Integration schemes for $\Delta b_{2}(\mathbf{r})$ and $\Delta\tau_{2}$.
Planar case corresponds to both (a) and (b). Spherical and cylindrical
cases correspond to (c) and (d). Diagram (c) is for fluids outside
of the region enclosed by the curved surface, while diagram (d) is
for fluids inside this region. Note that (c) and (d) in the limit
$R\rightarrow\infty$ corresponds to (a) and (b), respectively, which
are equivalent.\label{fig:IntegrationScheme}}
\end{figure}

The first non-trivial cluster terms are the ones of second order.
They describe the physical behavior of the inhomogeneous low density
gases up to order two in $z$. Our study of order terms use and generalize
ideas and analytic procedures taken from Refs. \cite{Bellemans_1962,Stecki_1980}.
Thus, we turn to one body density distribution, its second order residual
term is
\begin{equation}
\overline{\left\llbracket \rho^{(2)}(\mathbf{r}_{1})\right\rrbracket }\equiv g\bigl(\mathbf{r}_{1}\bigr)\int g\bigl(\mathbf{r}_{2}\bigr)f\bigl(\mathbf{r}_{12}\bigr)d\mathbf{r}_{2}\:.\label{eq:rho1IR}
\end{equation}
Obtaining $\rho(\mathbf{r})$ up to order $z^{2}$ is reduced to solving
this integral. In what follows, to proceed in the evaluation of order
two cluster integrals, we gradually introduce some conditions on the
system. We consider a system of spherical particles that interact
through an spherically symmetric pair potential, and then, the Mayer
function only depends on the distance between particles $r=\bigl|\mathbf{r}_{2}-\mathbf{r}_{1}\bigr|$.
We focus on the important case where $g(\mathbf{r})=1$ if $\mathbf{r}\in\mathcal{A}$,
a region bounded by the surface $\partial\mathcal{A}$, and is zero
otherwise. Therefore $Z_{1}$ coincides with $V$, the volume of
$\mathcal{A}$. The integrand of Eq. (\ref{eq:rho1IR}) can be written
as $f(r)+[g(\mathbf{r}_{2})-1]\,f(r)$ and 
\begin{eqnarray}
\int\!g\bigl(\mathbf{r}_{2}\bigr)f\bigl(r\bigr)d\mathbf{r}_{2} & = & 2b_{2}-\!\int\!\left[1-g\bigl(\mathbf{r}_{2}\bigr)\right]f\bigl(r\bigr)d\mathbf{r}_{2}\:.\label{eq:intrho01}
\end{eqnarray}
Here $1-g(\mathbf{r})=1$ if $\mathbf{r}\in\bar{\mathcal{A}}$ (where
we introduce the complement of a set $\bar{\mathcal{A}}=\mathcal{A}\setminus\mathbb{R}^{3}$)
and is zero otherwise. It is interesting to note that while the left
hand side in Eq. (\ref{eq:intrho01}) is $\int_{\mathcal{A}}f(r)d\mathbf{r}_{2}$
the integral on the right is $\int_{\bar{\mathcal{A}}}f(r)d\mathbf{r}_{2}$.
Turning to Eq. (\ref{eq:rho1IR}) we note that the term $g(\mathbf{r}_{1})\int_{\bar{\mathcal{A}}}f(r)d\mathbf{r}_{2}$
is non-null only if $\mathbf{r}_{1}$ is in the neighborhood of $\partial\mathcal{A}$
because $\mathbf{r}_{1}\in\mathcal{A}$, $\mathbf{r}_{2}\in\bar{\mathcal{A}}$,
and $r$ should be small enough to obtain $f(r)\neq0$. Thus, this
integral scales with the area of $\partial\mathcal{A}$. Now we restrict
further analysis to cases where surface $\partial\mathcal{A}$ has
constant curvature. We change integration variable $\mathbf{r}_{2}$
to relative coordinate between particles and introduce $u$ as the
distance between $\mathbf{r}_{1}$ and $\partial\mathcal{A}$. For
$\mathbf{r}_{1}\in\mathcal{A}$ we obtain 
\begin{equation}
2\Delta b_{2}\equiv-\int_{\bar{\mathcal{A}}}f(r)d\mathbf{r}_{2}=-\int_{\bar{\mathcal{A}}}S(r,u)f(r)dr\:.\label{eq:2Db20}
\end{equation}
Here, $S(r,u)$ is the surface area of an spherical shell with radius
$r$ and center $\mathbf{r}_{1}\in\mathcal{A}$ (at distance $u$
from $\partial\mathcal{A}$) that lies outside of $\mathcal{A}$.
In Fig. \ref{fig:IntegrationScheme} we give some insight about geometry-related
magnitudes for some simple shapes of $\partial\mathcal{A}$. The final
result is thus
\begin{equation}
\left\llbracket \rho^{(2)}(\mathbf{r}_{1})\right\rrbracket =\frac{2}{Z_{2}}g\bigl(\mathbf{r}_{1}\bigr)\left[2b_{2}+2\Delta b_{2}(u)\right]\:,\label{eq:rho1IR2}
\end{equation}
\begin{equation}
2\Delta b_{2}(u)=-\int_{r_{min}(u)}^{r_{max}(u)}S(r,u)f(r)dr\:,\label{eq:2Db2}
\end{equation}
where the boundary of the integration domain is explicit.

Now we consider $\tau_{2}$, given by
\begin{equation}
\tau_{2}\equiv\iint g(\mathbf{r}_{1})g(\mathbf{r}_{2})f\bigl(\mathbf{r}_{12}\bigr)d\mathbf{r}_{1}d\mathbf{r}_{2}\:.\label{eq:Tau2}
\end{equation}
and follow an approach similar to that used to transform Eq. (\ref{eq:rho1IR}).
For spherically symmetric pair potentials and a region $\mathcal{A}$
with arbitrary shape we found
\begin{eqnarray}
\tau_{2} & = & 2Z_{1}b_{2}+\Delta\tau_{2}\:,\nonumber \\
\Delta\tau_{2} & = & -\int_{\mathcal{A}}\left[\int_{\bar{\mathcal{A}}}f\bigl(r\bigr)d\mathbf{r}_{2}\right]d\mathbf{r}_{1}\:.\label{eq:dTau2}
\end{eqnarray}
Remarkably, $\Delta\tau_{2}$ is invariant under the interchange $\bar{\mathcal{A}}\leftrightarrow\mathcal{A}$
no matter the details of the potential between particles. It implies
that the inhomogeneous contribution to $\tau_{2}$ is the same if
the system is confined in a box or if it is confined to all the space
outside the box, ignoring of the shape of the box. This in-out symmetry
has been previously studied.\cite{Urrutia_2008,Urrutia_2010b} It
has a second interesting physical implication: if giving a certain
shape of $\mathcal{A}$ some terms of $\Delta\tau_{2}$ should change
its sign under the interchange $\bar{\mathcal{A}}\leftrightarrow\mathcal{A}$
all of them must be identically zero. For the case of a surface $\partial\mathcal{A}$
with constant curvature we transform integration variables $\mathbf{r}_{1}$
and $\mathbf{r}_{2}$, to position with respect to $\partial\mathcal{A}$
and relative coordinate between particles. Once the trivial integrations
are done one finds 
\begin{equation}
\Delta\tau_{2}=-\int_{\mathcal{A}}S(u)\left[\int_{\bar{\mathcal{A}}}S(r,u)f\bigl(r\bigr)dr\right]du\:.\label{eq:dTau20}
\end{equation}
Here, $S(u)$ is the area of the surface parallel to $\partial\mathcal{A}$
at a distance $u$ that lies in $\mathcal{A}$. Fig. \ref{fig:IntegrationScheme}
shows a picture of the overall approach for some simple shapes of
$\partial\mathcal{A}$. Expressions for $S(r,u)$ are known for planar,
spherical and cylindrical shapes of $\partial\mathcal{A}$ which shows
that Eq. (\ref{eq:dTau20}) is an interesting formula to analytically
evaluate $\tau_{2}$. Finally, one can introduce the boundary of the
integration domain to obtain the following two equivalent expressions
\begin{equation}
\Delta\tau_{2}=-\int_{u_{min}}^{u_{max}}S(u)\left[\int_{r_{min}(u)}^{r_{max}(u)}S(r,u)f(r)dr\right]du\:\label{eq:Tau2A1}
\end{equation}
{[}where the term in brackets is $2\Delta b_{2}(u)${]} and
\begin{equation}
\Delta\tau_{2}=-\int_{r_{min}}^{r_{max}}f(r)\left[\int_{u_{min}(r)}^{u_{max}(r)}S(u)S(r,u)du\right]dr\:.\label{eq:Tau2A2}
\end{equation}
Both expressions enable to evaluate $\Delta\tau_{2}$ for very simple
potentials like that of HS. Even though, Eq. (\ref{eq:Tau2A2}) condensates
the geometrical constraint in the inner integral over $u$ while the
nature of the interaction remains in $f(r)$. Thus, Eq. (\ref{eq:Tau2A2})
is a convenient starting point to analyze a variety of \emph{not so
simple} pair potentials. Next paragraphs introduce further simplifications
on Eqs. (\ref{eq:2Db2}) and (\ref{eq:Tau2A2}) for planar, spherical
and cylindrical confinements.

\subsection*{For planar walls\label{sub:PlanarWalls}}

In the planar case $S(r,u)=2\pi r(r-u)$, $r(u)_{min}=u$, $r_{max}(u)=\infty$
and Eq. (\ref{eq:2Db2}) takes the form

\begin{equation}
2\Delta b_{2}(u)=-2\pi\int_{u}^{\infty}(r^{2}-u\,r)f(r)dr\:.\label{eq:Db2pla}
\end{equation}
Obviously, $S(u)=A$, which corresponds to the area of an infinite
plane or the finite area of the hard plane in the unit cell when periodic
boundary conditions are used. Although, $r_{min}=0$, $r_{max}=\infty$,
$u_{min}(r)=0$ and $u_{max}(r)=r$. Therefore, Eq. (\ref{eq:Tau2A2})
reduces to
\begin{equation}
\Delta\tau_{2}=-2Aa_{2}\textrm{, with }a_{2}=\frac{\pi}{2}\int_{0}^{\infty}f(r)r^{3}dr\:.\label{eq:DTau2pla}
\end{equation}
It is interesting to compare Eq. (\ref{eq:DTau2pla}) with the bulk
second cluster integral of a system in a region of volume $V$, 
\begin{equation}
\tau_{2}=2Vb_{2}\textrm{, with }b_{2}=2\pi\int_{0}^{\infty}f(r)r^{2}dr\:,\label{eq:Tau2Bulk}
\end{equation}
that gives the bulk system second virial coefficient $B_{2}=-b_{2}$.
Thus we show that the difficult of solving the terms $2\Delta b_{2}(u)$
and $\Delta\tau_{2}$ for an inhomogeneous fluid confined by a planar
wall is similar to the one of solving the bulk fluid term $\tau_{2}$.
Both $\Delta b_{2}$ and $\Delta\tau_{2}$ are known for the HS and
the SW systems.\cite{Bellemans_1962_b,Stecki_1980}

\subsection*{For spherical walls\label{sub:SpherWalls}}

Again, we start from Eq. (\ref{eq:Tau2A2}). Two different situations
arise because the system may be inside the sphere with area $A=4\pi R^{2}$
and volume $V=4\pi R^{3}/3$ or outside of it. If the fluid is outside
of the spherical surface $\partial\mathcal{A}$ then $u:(0,\infty)$,
$S(u)=4\pi(R+u)^{2}$ and $S(r,u)=\pi r(r-u)(2R+u-r)/(R+u)$. On the
opposite, for fluids inside the spherical shell $\partial\mathcal{A}$
one finds $u:(0,R)$, $S(u)=4\pi(R-u)^{2}$ and, if $u<r<2R-u$ then
$S(r,u)=\pi r(r-u)(2R-u+r)/(R-u)$ but if $r>2R-u$ then $S(r,u)=4\pi r^{2}$.
For a geometrical insight see Fig. \ref{fig:IntegrationScheme} (c)
and (d).

In the case of a fluid that surrounds an spherical object we have
\begin{equation}
2\Delta b_{2}(u)=\frac{\pi}{R+u}\int_{u}^{2R+u}\!\!\!\!f(r)r(u-r)(2R+u-r)dr\:,\label{eq:Db2sphOUT}
\end{equation}
and if the fluid is inside of an spherical cavity we have:
\begin{eqnarray}
2\Delta b_{2}(u) & = & \frac{\pi}{R-u}\int_{u}^{2R-u}\!\!\!\!f(r)r(u-r)(2R-u+r)dr-\nonumber \\
 &  & 4\pi\int_{2R-u}^{\infty}f(r)r^{2}dr\:.\label{eq:Db2sphIN}
\end{eqnarray}
To evaluate $\Delta\tau_{2}$ one may assume that the fluid is outside
of an spherical shell to obtain
\begin{equation}
\Delta\tau_{2}=-\int_{0}^{\infty}f(r)w(r)dr\:,\label{eq:DTau2sph}
\end{equation}
with $w(r)=A\pi r^{3}-\frac{1}{3}\pi^{2}r^{5}$ for $0<r\leq2R$ while
$w(r)=V4\pi r^{2}$ for $r>2R$. We have verified that Eq. (\ref{eq:DTau2sph})
{[}with the same expression for $w(r)${]} also applies to the case
of fluid inside of a spherical shell. The situation of small curvature
is simpler to analyze by writing
\begin{equation}
\Delta\tau_{2}=-2Aa_{2}+2c_{2}+2d_{2}\:,\label{eq:DTau2sphRlarge}
\end{equation}
where $a_{2}$ was defined in Eq. (\ref{eq:DTau2pla}) and 
\begin{eqnarray}
c_{2} & = & \frac{\pi^{2}}{6}\int_{0}^{\infty}f(r)r^{5}dr\:,\label{eq:c2def}\\
d_{2} & = & \int_{2R}^{\infty}f(r)\left(-V2\pi r^{2}+A\frac{\pi}{2}r^{3}-\frac{\pi^{2}}{6}r^{5}\right)dr\:,\label{eq:d2def}
\end{eqnarray}
those make sense if $c_{2}$ converges. Note that $a_{2}$ and $c_{2}$
depend on temperature but not on $R$. Although, $d_{2}$ in Eq. (\ref{eq:DTau2sphRlarge})
may include terms $\mathcal{O}(R^{0})$ and higher order ones in $R^{-1}$.
If $c_{2}$ does not converge, it is preferable to define a single
term $\tilde{c}_{2}=c_{2}+d_{2}$ as 
\begin{equation}
\tilde{c}_{2}=\frac{\pi^{2}}{6}\!\!\int_{0}^{2R}\!\!\!f(r)r^{5}dr-\!\int_{2R}^{\infty}\!\!f(r)(V2\pi r^{2}-A\frac{\pi}{2}r^{3})dr\,.\label{eq:c2m}
\end{equation}
Note that Eqs. (\ref{eq:DTau2sphRlarge}-\ref{eq:c2m}) apply to
both, finite values of $R$ and the case of small curvature $R\gg1$.

Interestingly, if the pair potential is of finite range smaller than
$2R$, integral $c_{2}$ converges, $d_{2}=0$, $a_{2}$ and $c_{2}$
are functions of $T$ (do not depend on $R$), and 
\begin{equation}
\Delta\tau_{2}=-2Aa_{2}+2c_{2}\:,\label{eq:DTau2sphFinRange}
\end{equation}
without the term of order $R$. For example, this is the case of the
truncated 12-6 Lennard Jones potential (it does not matter if it is
shifted or not) which is frequently used in simulations and theoretical
development.\cite{Allen1987,Shaul_2010} Naturally, it is also the
case of the hard sphere (HS) and square well (SW) interactions. For
HS and SW fluids one finds 
\begin{equation}
b_{2}=-\frac{2\pi}{3}\:,\:\:a_{2}=-\frac{\pi}{8}\:,\:\:c_{2}=-\frac{\pi^{2}}{36}\:,\label{eq:bac2HS}
\end{equation}
\begin{equation}
b_{2}=\frac{-2\pi}{3}\Sigma(3)\,,\:a_{2}=-\frac{\pi}{8}\Sigma(4)\,,\:c_{2}=-\frac{\pi^{2}}{36}\Sigma(6)\:,\label{eq:bac2SW}
\end{equation}
respectively (here results are given in units of the hard-core diameter
$\sigma$, $\Sigma(x)=\lambda^{x}+e^{-\beta\epsilon}(1-\lambda^{x})$,
being $\lambda$ the range of the square well and $\epsilon$ its
depth). These results are consistent with those found previously using
different approaches\cite{Bellemans_1963,McQuarrie_1987,Urrutia_2010b,Urrutia_2011}\footnote{We have found a typo in $a_{2}$ and $c_{2}$ taken from Ref. \cite{Urrutia_2011}.
It was amended here in the results of $a_{2}$ and $c_{2}$ for square-well
interaction. } and serve here to cross-check new expressions.

\subsection*{For cylindrical walls\label{sub:CylindWalls}}

In this geometry the system may be inside the cylinder of area $A=2\pi R\,L$
and volume $V=\pi R^{2}L$ or outside of it. Now, if the fluid is
outside of the cylindrical surface $\partial\mathcal{A}$ then, $S(u)=2\pi\left(R+u\right)L$,
else, if it is inside the cylindrical wall then $S(u)=2\pi\left(R-u\right)L$.
Note that in this case $L$ is the length of an infinite cylinder
or a finite length when periodic boundary conditions are used. Analytic
expression of $S(r,u)$ involves elliptic integrals of the first,
second and third kind\cite{Lamarche_1990} and are given in the Appendix
\ref{Apsec:CylWalls}. To evaluate $\Delta\tau_{2}$ one may assume
again that the fluid is outside of the cylindrical shell and obtain
a relation identical to Eq. (\ref{eq:DTau2sph}). However, for the
cylindrical confinement we do not find a simple analytic expression
for $w$. For large $R$ the series expansion of $S(r,u)$ provides
the expression 
\begin{equation}
w(r)=A\pi r^{3}-L\left(\frac{\pi^{2}r^{5}}{16R}+\frac{\pi^{2}r^{7}}{512R^{3}}\right)+\mathcal{O}(r^{9}R^{-5})\:,\label{eq:wcyl}
\end{equation}
which applies to the region $0<r<2R$.

Thus, we obtain
\begin{equation}
\Delta\tau_{2}=-2Aa_{2}+2x_{2}+2d_{2}\:,\label{eq:DTau2cylRlarge}
\end{equation}
\begin{equation}
x_{2}=\frac{L}{R}\frac{3}{16}c_{2}\:,\label{eq:x2c2}
\end{equation}
 {[}$a_{2}$ and $c_{2}$ are given by Eqs. (\ref{eq:DTau2pla}, \ref{eq:c2def}){]}
and 
\begin{eqnarray}
d_{2} & = & \int_{2R}^{\infty}f(r)\left(-V2\pi r^{2}+A\frac{\pi}{2}r^{3}-\frac{L}{R}\frac{\pi^{2}}{32}r^{5}\right)dr+\nonumber \\
 &  & L\int_{0}^{2R}f(r)\left[\frac{\pi^{2}r^{7}}{1024R^{3}}+\mathcal{O}\left(\frac{r^{9}}{R^{5}}\right)\right]dr\:.\label{eq:d2defcyl}
\end{eqnarray}
Naturally this makes sense if $x_{2}$ (i.e. $c_{2}$) converges.
Otherwise, it is convenient to define a single term $\tilde{x}_{2}=x_{2}+d_{2}$
following the same criteria adopted in Eq. (\ref{eq:c2m}). When the
pair potential has finite range smaller than $2R$ the convergence
of $c_{2}$ is secured, first term of $d_{2}$ is null but higher
order terms in $R^{-1}$ do not disappear in the cylindrical case.
We found for the HS and SW particles,
\begin{equation}
x_{2}=-\frac{\pi^{2}}{192}\textrm{ and }x_{2}=-\frac{\pi^{2}}{192}\Sigma(6)\:.\label{eq:x2HS-SW}
\end{equation}
The HS result was found before utilizing a different method but SW
result is new.\cite{Urrutia_2010b}

\section{Case study: $\tau_{2}$ for the confined Lennard-Jones fluid\label{sec:Apps}}

Virial series in general and specifically its truncation at second
order coefficient $B_{2}$ have been thoroughly studied for a long
time because they enable to analytically describe the properties of
diluted homogeneous fluids. Beyond the case of HS and SW potentials,
analytic expressions of $B_{2}(T)$ were found for the 12-6 Lennard-Jones
(LJ) potential\cite{Vargas_2001}, for the $2k$-$k$ LJ potential\cite{Glasser_2002}
and others LJ-like potentials.\cite{GonzalezCalderon_2015} For molecular
dynamic simulation purposes the truncation of interaction potential
at finite range is necessary. Yet, virial coefficients of truncated-LJ
systems were numerically evaluated.\cite{Shaul_2010}

Virial series are not a standard method to study inhomogeneous fluids.
Nonetheless, a few recent works studied the inhomogeneous HS fluid
under this framework.\cite{Yang_2013,Urrutia_2014,Yang_2015} In the
case of inhomogeneous LJ fluid we found a single work that based on
this series (truncated at second order) study the adsorption of 12-6
LJ gas on a planar attractive wall.\cite{Stecki_1980} There, second
order cluster integral is numerically evaluated. In the following,
cluster integral $\tau_{2}$ for the inhomogeneous 2k-k LJ confined
by hard walls of constant curvature is evaluated analytically for
the first time.

\subsection{$\tau_{2}$ for the inhomogeneous Lennard-Jones fluid\label{sub:Inhom-LJ}}

Here we evaluate analytically $\tau_{2}$ by applying to Eqs. (\ref{eq:Db2pla})
and (\ref{eq:DTau2pla}) some ideas and procedures partially taken
from Refs. \cite{Vargas_2001,Glasser_2002}. We generalize those calculus
to obtain the second cluster integral for planar spherical, and cylindrical
confinement.

One can observe that several of the integrals appearing in Sec. \ref{sec:SndOrderTerms}
are of the form $\int_{a_{1}}^{a_{2}}f(r)r^{m}dr$. For $m=2$ (and
$a_{1}=0$, $a_{2}\rightarrow\infty$) it corresponds to $b_{2}$
and $B_{2}$ that describe homogeneous systems {[}see Eq. (\ref{eq:Tau2Bulk}){]}.
We introduce the $2k$-$k$ LJ pair potential
\begin{equation}
\phi(r)=4\epsilon\left[\left(\frac{\sigma}{r}\right)^{2k}-\left(\frac{\sigma}{r}\right)^{k}\right]\:,\label{eq:LJpot}
\end{equation}
with $k\geq6$. The case $k=6$ is the most used to model simple monoatomic
fluids, yet higher values like $k=18$ are utilized in studies of
particles with short range interaction potential as neutral colloids.\cite{Vliegenthart_2000}
Thus, we shall solve integrals of the type
\begin{equation}
\int\sigma^{-(m+1)}\left\{ \exp\left[-\beta\phi(r)\right]-1\right\} r^{m}dr\:.\label{eq:C0}
\end{equation}
with $0<m+1\leq k$. In the case of Eqs. (\ref{eq:DTau2pla}, \ref{eq:Tau2Bulk})
and (\ref{eq:c2def}) the integration domain $(0,\infty)$ leave us
with indefinite integrals. Changing variable to $x=r/\sigma$ and
defining $z=4\beta\epsilon$ one finds
\begin{equation}
C_{m+1,k}=\int_{l_{1}}^{l_{2}}\left\{ \exp\left[-z\left(x^{-2k}-x^{-k}\right)\right]-1\right\} x^{m}dx\:,\label{eq:Cx}
\end{equation}
where $l_{2}$ is typically $\sigma/2R$ or $\infty$. Changing variables
to $y=x^{k}$ it transforms to
\begin{equation}
C_{m+1,k}=\frac{1}{k}\int_{l_{1}^{k}}^{l_{2}^{k}}y^{q-1}\left\{ \exp\left[-z\left(y^{-2}-y^{-1}\right)\right]-1\right\} dy\:.\label{eq:Cy}
\end{equation}
where $q=\frac{m+1}{k}$. A comparison between Eqs. (\ref{eq:Cx})
and (\ref{eq:Cy}) shows that $C_{m+1,k}=\frac{1}{k}C_{q,1}$. We
transform variables to $u=y^{-1}$ and fix $l_{1}^{-k}=M$ (i.e. $l_{1}=M^{-1/k}$)
and $l_{2}^{-k}=\varepsilon$ to obtain 
\begin{equation}
C_{q,1}=\int_{\varepsilon}^{M}u^{-(q+1)}\left\{ \exp\left[-z\left(u^{2}-u\right)\right]-1\right\} du\:.\label{eq:Cu}
\end{equation}
It is convenient to define $C_{q}(\varepsilon)=\underset{M\rightarrow\infty}{\lim}C_{q,1}$
to analyze the condition $l_{1}=0$ and thus we assume $q>0$ to prevent
the divergence. Once we integrate Eq. (\ref{eq:Cu}) by parts we obtain
\begin{equation}
qC_{q}(\varepsilon)=\varepsilon^{-q}\!\left\{ \exp\!\left[-z\!\left(\varepsilon^{2}-\varepsilon\right)\right]\!-\!1\right\} +I_{q,\varepsilon}-2I_{q-1,\varepsilon}\label{eq:Cuq}
\end{equation}
where $I_{\nu,\varepsilon}=z\int_{\varepsilon}^{\infty}u^{-\nu}\exp\left[-z\left(u^{2}-u\right)\right]du$.
$I_{\nu}=I_{\nu,0}$ and $C_{q}(0)$ were studied by Glasser \cite{Glasser_2002}
who gives closed expressions for $\nu<1$ and $0<q<1$, respectively,
in terms of Kummer's hypergeometric functions. In Appendix \ref{Apsec:InuyC}
we analyze the functions $I_{\nu,\varepsilon}$ and $C_{q}(\varepsilon)$,
and provide explicit expressions of them when $0<\varepsilon\ll1$.

Before analyzing the asymptotic behavior at large radius we give in
terms of $C_{q}$ the following exact expressions
\begin{equation}
\frac{\tau_{2}}{2}=V\frac{2\pi}{k}C_{3/k}(0)-A\frac{\pi}{2k}C_{4/k}(0)\:,\label{eq:plane}
\end{equation}
\begin{equation}
\left.\frac{\tau_{2}}{2}\right|_{\textrm{in}}=V\frac{2\pi}{k}C_{3/k}(\varepsilon)-A\frac{\pi}{2k}C_{4/k}(\varepsilon)+\frac{\pi^{2}}{6k}C_{6/k}(\varepsilon)\:,\label{eq:Tau2sphe}
\end{equation}
which apply to the planar and spherical cases, respectively. In Eq.
(\ref{eq:Tau2sphe}) and from now on we fix $\varepsilon=\left(2R\right)^{-k}$
($\sigma$ is the unit length). For the cylindrical case we found
\begin{eqnarray}
\left.\frac{\tau_{2}}{2}\right|_{\textrm{in}} & = & V\frac{2\pi}{k}C_{3/k}(\varepsilon)-A\frac{\pi}{2k}C_{4/k}(\varepsilon)+\nonumber \\
 &  & \frac{L}{R}\frac{\pi^{2}}{32k}C_{6/k}(\varepsilon)+\frac{L}{R^{3}}\frac{\pi^{2}}{512k}C_{8/k}(\varepsilon)+....\:,\label{eq:Tau2cyle}
\end{eqnarray}
where higher order $C_{q}$ functions were neglected. In the limit
of large $R$ ($\varepsilon\rightarrow0$) we found the following
expressions for $B_{2}=-b_{2}$, $b_{2}$, $a_{2}$, $c_{2}$, $d_{2}$
and $\tilde{c}_{2}$:
\begin{eqnarray}
b_{2} & = & \frac{2\pi}{k}C_{3/k}(0)\:,\label{eq:b2}\\
a_{2} & = & \frac{\pi}{2k}C_{4/k}(0)\:,\label{eq:a2}\\
c_{2} & = & \frac{\pi^{2}}{6k}C_{6/k}(0)\:,\label{eq:c2}
\end{eqnarray}
that are given in $\sigma$ units. Explicit form of $C_{q}(0)$ in
terms of hypergeometric functions is given in Appendix \ref{Apsec:InuyC}
Eq. (\ref{aeq:Cq0}). Note that Eq. (\ref{eq:c2}) applies if $k>6$,
when $c_{2}$ converges. In this case $d_{2}=-\frac{V2\pi}{k}\Delta C_{3/k}+\frac{A\pi}{2k}\Delta C_{4/k}-\frac{\pi^{2}}{6k}\Delta C_{6/k}$
with $\Delta C_{q}=C_{q}(0)-C_{q}(\varepsilon)$. From the series
expansion we obtain
\begin{equation}
d_{2}=-(2R)^{6-k}\frac{4\pi^{2}}{T(k-6)(k-4)(k-3)}+\mathcal{O}(R^{6-2k})\:.\label{eq:d2}
\end{equation}
The relevant case $k=6$ corresponds to the 12-6 LJ potential which
produces $\tilde{c}_{2}=-\frac{V\pi}{3}\Delta C_{1/2}+\frac{A\pi}{12}\Delta C_{2/3}+\frac{\pi^{2}}{36}C_{1}(\varepsilon)$.
It can be split in 
\begin{equation}
c_{2}=\frac{\pi^{2}}{18T}\left[12\ln\left(2R\right)+\ln\left(\frac{T}{4}\right)+10\right]+\frac{\pi^{2}}{36}reg[C_{1}(0)]\:,\label{eq:c2uno}
\end{equation}
\begin{equation}
d_{2}=(2R)^{-6}\frac{\pi^{2}(T-2)}{108T^{2}}+\mathcal{O}(R^{-12})\:,\label{eq:d2uno}
\end{equation}
where $reg[C_{1}(0)]$ is the regular (non-divergent) part of $C_{1}(\varepsilon)$
in the limit $\varepsilon\rightarrow0$. For cylindrical walls the
details of the calculus are given in the Appendix \ref{Apsec:CoeffCyl}.
Here we show the main results: if $k>6$ then $x_{2}=\frac{L}{R}\frac{3}{16}\frac{\pi^{2}}{6k}C_{6/k}(0)$
{[}from Eqs. (\ref{eq:x2c2}) and (\ref{eq:c2}){]} else, if $k=6$
then
\begin{equation}
x_{2}=\frac{\pi^{2}L}{92TR}\left[12\ln\left(2R\right)+\ln\left(\frac{T}{4}\right)\right]+\mathcal{O}\left(\frac{L}{R}\right)\:.\label{eq:x2}
\end{equation}

We found that $X=b_{2},a_{2},c_{2},x_{2}$ for large $k$ values
behaves as it was made of hard spheres, i.e. $\underset{k\rightarrow\infty}{\lim}X=X_{\textrm{HS}}$
with $X_{\textrm{HS}}$ the coefficients of the HS confined system
described by Eqs. (\ref{eq:bac2HS}) and (\ref{eq:x2HS-SW}). This
checks the overall consistence of our results.

In thermodynamic perturbation theories it is required to obtain the
effective particle diameter of a fluid. For LJ fluids this effective
diameter is also related with $C_{q}$. Barker and Henderson had given
two possible definition that are widely used in the literature. The
hard-core reference corresponds to $\sigma_{\textrm{eff}}=-\frac{1}{k}C_{1/k}(1)$,\cite{Barker_1967_b}
while that adopted on soft-core reference systems is $\sigma_{\textrm{eff}}=-\frac{1}{k}C_{1/k}(0)$.\cite{HendersonD_1970,Hansen2006}
Barker and Henderson proposals are used to study fluids systems using
a variety of techniques including density functional theories and
the law of corresponding states.\cite{Tang_2003,Orea_2015} 

We also calculated $\Delta b_{2}(0)=\underset{u\rightarrow0}{\lim}\Delta b_{2}(u)$
related with the contact- or wall-density $\rho_{\textrm{c}}=\underset{u\rightarrow0}{\lim}\rho(u)$
through Eq. (\ref{eq:rho1IR2}). We found 
\begin{eqnarray}
\Delta b_{2}(0) & = & -\frac{\pi}{k}C_{3/k}(0)\:,\label{eq:Db20pl}\\
\Delta b_{2}(0) & = & -\frac{\pi}{k}C_{3/k}(\varepsilon)\pm\frac{\pi}{2Rk}C_{4/k}(\varepsilon)\:,\label{eq:Db20sph}
\end{eqnarray}
for planar and spherical cases respectively (plus sign corresponds
to fluid surrounding the shell and the minus sign to the opposite
case). We note that for $k\geq6$ Eq. (\ref{eq:Db20sph}) does not
include term $C_{1}(\varepsilon)$ and thus logarithmic dependence
is absent from $\Delta b_{2}(0)$. The expansion of Eq. (\ref{eq:Db20sph})
produces 

\begin{equation}
\Delta b_{2}(0)=-\frac{\pi}{k}C_{3/k}(0)\pm\frac{\pi}{k2R}C_{4/k}(0)+\mathcal{O}(R^{3-k})\:.\label{eq:Db20R}
\end{equation}

We point out that our approach is directly extendable to systems
with dimension $d\neq3$, e.g. $d=2$. During decades, several works
aimed to study two-dimensional fluids composed by particles with hard-core
interaction (the so called hard discs) and also LJ potential.\cite{Ashwin_2009,Khordad_2012}
In the case of a planar wall that cut the $d$-space in two equal
regions (one of which is available for particles), one should replace
in Eq. (\ref{eq:C0}) $m$ by $d-1+m'$, $m'=0$ corresponds to the
bulk $b_{2}$ and $m'=1$ corresponds to the planar term $a_{2}$.
For a $d$-spherical wall one finds that term of order $R^{d-2}$
($m'=2$) is zero and $m'=3$ corresponds to $c_{2}$ (order $R^{d-3}$).
Expressions of $S(u,r)$, which measures the volume of overlap between
two $d$-spheres, were given in Ref. \cite{Urrutia_2010}.
\begin{table}
\begin{centering}
\begin{tabular}{|c|c|c|c|c|c|c|}
\hline 
k & $6$ & $7$ & $8$ & $9$ & $12$ & $18$\tabularnewline
\hline 
\hline 
$b_{2}$
 & $3.418$ & $2.412$ & $1.883$ & $1.560$ & $1.074$ & $0.727$\tabularnewline
\hline 
$a_{2}$
 & $9.016$ & $5.003$ & $3.418$ & $2.601$ & $1.560$ & $0.946$\tabularnewline
\hline 
$c_{2}$, $x_{2}$ & $-$ & $56.83$ & $17.15$ & $9.016$ & $3.418$ & $1.560$\tabularnewline
\hline 
$T_{c}$ & $1.312$ & $0.997$ & $0.831$ & $0.730$ & $0.560$ & $0.425$\tabularnewline
\hline 
\end{tabular}
\par\end{centering}

\caption{Boyle temperature for each coefficient at different values of $k$.
As a reference in the temperature scale we include the critical temperature
$T_{c}$, taken from Ref.\cite{Vliegenthart_1999}, except for k$=6$
taken from Ref. \cite{PerezPellitero_2006}.\label{tab:BoyleTemp}}
\end{table}
One can inquire what makes evident the dependence of system properties
on its inhomogeneous nature. Thus, focusing in $\tau^{2}$, the stronger
confinement increases the ratio $A/V$. Also, temperature that makes
$b_{2}=0$ enables to enhance the presence of $a_{2}$, and $T$ value
that makes null $a_{2}$ enables to enhance the effect of $c_{2}$.
In Table \ref{tab:BoyleTemp} we present the Boyle temperature i.e.
$T$ values at which each of the first three coefficients of $\tau_{2}$,
i.e. $b_{2}$, $a_{2}$ and $c_{2}$, are zero. There the critical
temperature is also given for comparison. Given that $c_{2}$ for
$k=6$ depends on both $T$ and $R$ {[}see Eq. (\ref{eq:c2uno}){]}
there is not a unique value of $T$ at which $c_{2}=0$. 
\begin{figure}
\begin{centering}
\includegraphics[width=0.85\columnwidth]{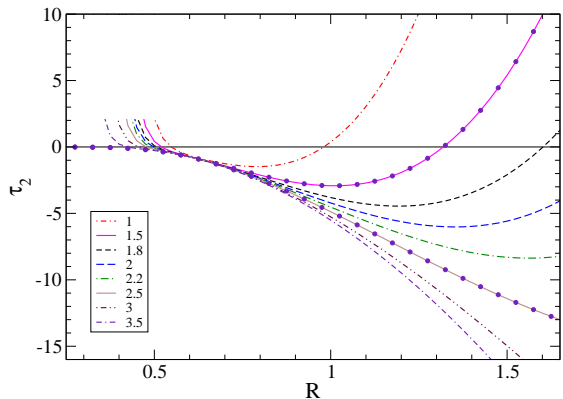}
\par\end{centering}

\caption{Second cluster integral $\tau_{2}$ for the 12-6 LJ system confined
by a spherical wall. Each curve corresponds to a different temperature.\label{fig:Tau2RT}}
\end{figure}
As an example of the obtained results in Fig. \ref{fig:Tau2RT} we
plot the dependence with $R$ of the second cluster integral for the
12-6 LJ fluid confined in a spherical pore. Curves show the asymptotic
expression for large $R$, including terms of order $\mathcal{O}(R^{-18})$
(order $\varepsilon^{3}$ in $d_{2}$). Dots show numerical evaluation
of the exact integral $\tau_{2}$. The highest considered $T$ is
near to the Boyle temperature for $b_{2}$ therefore in this case
$\tau_{2}$ is driven by $a_{2}A$. For very small radius exact results
smoothly goes to zero. Yet, for $R\lesssim0.6$ curves differ significantly
form the exact results and isotherms of smaller temperature separates
at higher radii from the exact result. We observed a similar behavior
for other values of $k>6$ which do not include the logarithmic dependence
with $R$.

\section{Results\label{sec:Results}}

Before analyzing the consequences on the LJ system of the obtained
expressions for $\tau_{2}$, it is interesting to discuss in the present
context some general relations known as exact sum rules. Once the
volume notion is introduced, a further step in the thermodynamic interpretation
of the confined system can be done. Starting from $\Omega$ we obtain
the pressure that the system makes on the surrounding walls, the pressure
on the wall $P_{\textrm{w}}$, that is the key magnitude of the reversible
work term in the first law of thermodynamics. For systems confined
by constant-curvature walls and using the volume notion defined in
Sec. \ref{sec:SndOrderTerms} $P_{\textrm{w}}$ is 
\begin{eqnarray}
P_{\textrm{w}} & = & -\frac{dR}{dV}\left.\frac{\partial\Omega}{\partial R}\right|_{\mu,T}\:.\label{eq:Pw}
\end{eqnarray}
The relation between $P_{\textrm{w}}$, $T$ and $z$ is the basic
EOS of these inhomogeneous systems. It is not necessary to deal with
bulk or surface properties. A second exact relation valid for constant-curvature
hard confinement is 
\begin{equation}
P_{\textrm{w}}=\rho_{\textrm{c}}T,\label{eq:ContactTh}
\end{equation}
which is a contact theorem with $\rho_{\textrm{c}}=\rho(r=R)$. Both
exact relations (\ref{eq:Pw}) and (\ref{eq:ContactTh}) are a convenient
starting point to analyze the system properties.

For the three geometrical constraints we decompose $\Omega$ in Eq.
(\ref{eq:gcO}) as 
\begin{equation}
\Omega=-PV+\gamma A\:,\label{eq:OmegaGam}
\end{equation}
with bulk pressure $P=-\left.\frac{\partial\Omega}{\partial V}\right|_{\mu,T,A,R}$
and fluid-substrate surface tension $\gamma=\left(\Omega+PV\right)/A$.
This definition identifies the fluid-substrate surface tension with
an excess of free energy (over bulk and per unit area), but it is
not the unique possible definition that can be adopted (for example
$\left.\frac{\partial(\Omega+PV)}{\partial A}\right|_{\mu,T,V}$ is
also sometimes used). Given that $\Omega$ and $P$ are the actual
exact grand free-energy of the system and pressure of the reservoir
at the same temperature and chemical potential, $\gamma$ defined
in Eq. (\ref{eq:OmegaGam}) strongly depends on the adopted measures
of volume and surface area to describe the system properties. Mapping
the results between different conventions may be done with a little
of linear algebra.\cite{Urrutia_2014}

Once the decomposition of $\Omega$ in Eq. (\ref{eq:OmegaGam}) is
assumed there is a third \emph{exact sum rule} that applies to spherical
and cylindrical confined system
\begin{eqnarray}
\Delta P & = & \gamma\frac{sc}{R}+\left.\frac{\partial\gamma}{\partial R}\right|_{\mu,T}\:,\label{eq:Lapl2}
\end{eqnarray}
where $\Delta P=\pm\left(P_{\textrm{w}}-P\right)$. Plus sign applies
to the case of fluid in the outer region while minus sign applies
to opposite. The parameter $sc$ is: $sc=2$ if the surface is a sphere,
$sc=1$ for the cylinder. To include planes one may consider $sc=0$.
Eq. (\ref{eq:Lapl2}) is the \emph{exact} form that takes the Laplace
equation in this context. In the case of the LJ fluid our expression
of $\tau_{2}(R)$ enables to analytically evaluate $\Omega$, $P_{\textrm{w}}$,
$\rho_{\textrm{c}}$ and $\gamma$ up to order $z^{2}$.

\subsection{Low density inhomogeneous gas\label{sub:LowDens}}

We consider the unconstrained open system {[}$m\rightarrow\infty$
in Eq. (\ref{eq:gcO}){]} of LJ particles at low density confined
by planar, spherical or cylindrical walls. Then, we truncate Eq. (\ref{eq:Omgz})
at second order to obtain $\beta\Omega=-zV-z^{2}\frac{1}{2}\tau_{2}$.
Therefore the first consequence of our calculus on $\tau_{2}$ is
that grand-free energy of 2$k$-$k$ LJ fluid contains the expected
terms linear with volume and surface area. These terms are identical
for the three studied geometries. At planar geometry, no extra term
exist as symmetry implies for all $\tau_{i}$. In case of spherical
confinement a term linear with total normal curvature of the surface
$A2/R\propto R$ does not appear at order $z^{2}$ but it should exist
at higher ones. A term linear with total Gaussian curvature $A/R^{2}\propto\textrm{constant}$
exist. Extra terms that scales with negative powers of $R$ were also
found. A logarithmic term proportional to $\ln R^{-1}$ was recognized
only for $k=6$. The cylindrical confinement is not very different
to the spherical case. We simply trace the differences: even that
Gaussian curvature is zero in this geometry, a term linear with $A/R^{2}\propto L/R$
was found. The existence of a logarithmic term for $k=6$ was verified,
in this case it was proportional to $L\ln R^{-1}/R$.

Up to order $z^{2}$ we found the series expansions 
\begin{eqnarray}
\beta\Omega & = & -(z+z^{2}b_{2})V-z^{2}\frac{1}{2}\Delta\tau_{2}\:,\nonumber \\
\beta P_{\textrm{w}} & = & z+z^{2}b_{2}+\frac{z^{2}}{2}A^{-1}\partial\Delta\tau_{2}/\partial R\:,\nonumber \\
\rho_{\textrm{c}} & = & z+z^{2}2b_{2}+z^{2}2\Delta b_{2}(u=0)\:.\label{eq:z2series}
\end{eqnarray}
Last two relations through Eq. (\ref{eq:ContactTh}) imply $2\Delta b_{2}(u=0)=-b_{2}+A^{-1}\partial\Delta\tau_{2}/\partial R$.
Furthermore, for bulk homogeneous system we found $\beta P=z+z^{2}b_{2}$
and $\rho_{\textrm{b}}=z+z^{2}2b_{2}$ (subscript b refers to the
bulk at the same $T$ and $\mu$ conditions).

For the low density LJ fluid we obtained\cite{Urrutia_2014} 
\begin{equation}
\beta\gamma=-\frac{\Delta\tau_{2}}{2A}z^{2}=-\frac{\Delta\tau_{2}}{2A}\rho_{\textrm{b}}^{2}\:,\label{eq:Gamrhob}
\end{equation}
that are exact up to $\mathcal{O}(z^{3})$ and $\mathcal{O}(\rho_{\textrm{b}}^{3})$.
For planar, spherical and cylindrical walls it reduces to 
\begin{equation}
\gamma=a_{2}T\rho_{\textrm{b}}^{2}\:,\:\:\gamma=\left(a_{2}-\frac{c_{2}+d_{2}}{4\pi R^{2}}\right)T\rho_{\textrm{b}}^{2}\:,\label{eq:gammaps}
\end{equation}
\begin{equation}
\gamma=\left(a_{2}-\frac{3}{32\pi R^{2}}c_{2}+...\right)T\rho_{\textrm{b}}^{2}\:,\label{eq:gammac}
\end{equation}
respectively. In Eq. (\ref{eq:gammac}) higher order terms are: $\mathcal{O}(R^{-4})$
for $k>6$, but includes terms $\mathcal{O}(R^{-2})$ if $k=6$ (even,
it is exact up to order $R^{-2}\ln R$). Thus, results of Eqs. (\ref{eq:a2})
to (\ref{eq:d2uno}) enable us for the first time to study on analytic
grounds the wall-fluid surface tension of the LJ systems for planar,
spherical and cylindrical walls, at low density. 
\begin{figure}
\centering{}\includegraphics[width=0.85\columnwidth]{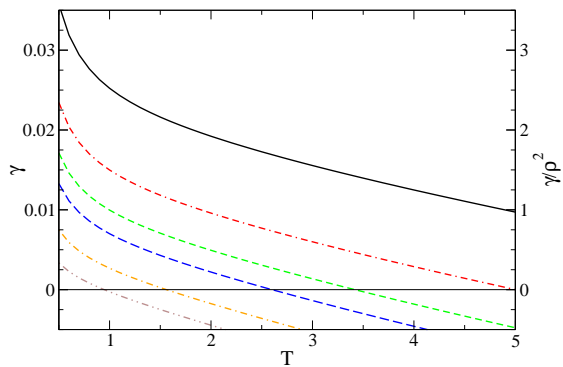}\caption{Surface tension of a 2$k$-$k$ LJ fluid in contact with a planar
wall, we fix $\rho_{\textrm{b}}=0.1$ and consider various $k$ values.
From top to bottom $k$ increases. Curves correspond to $k=6,7,8,9,12,18$.\label{fig:gammaPlanar}}
\end{figure}
In Fig.\ref{fig:gammaPlanar} it is shown the surface tension of the
2$k$-$k$ LJ gas confined by a planar wall for different values of
$k$. All cases show a monotonous decreasing behavior of $\gamma$
with $T$. At low temperatures $\gamma$ is positive (it diverges
as $\exp(1/T)$ as $T\rightarrow0$) and becomes negative at high
temperatures. The temperature where $\gamma$ is zero is lower for
bigger $k$ (temperatures are given in Tab. \ref{tab:BoyleTemp},
second row). In the case of the 12-6 LJ system we found $\gamma\approx0.035$
at $T=0.5$ and $\gamma\approx0.01$ at $T=5$ ($\rho_{\textrm{b}}=0.1$).
Scale on the right shows $\gamma/\rho_{\textrm{b}}^{2}$ which is
independent of density. 
\begin{figure}
\centering{}\includegraphics[width=0.85\columnwidth]{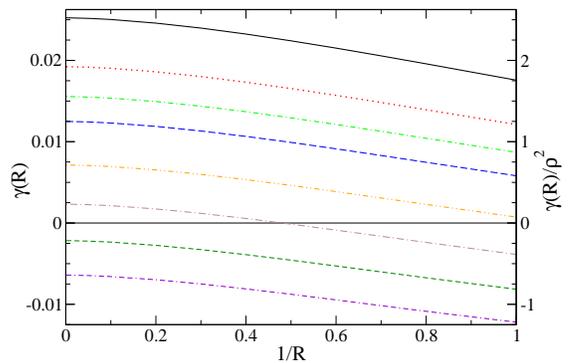}\caption{Surface tension of the 12-6 LJ fluid in contact with a spherical wall
at $\rho_{\textrm{b}}=0.1$ (for both concave and convex shapes) and
at various temperatures. From top to bottom $T=1,2,3,4,6,8,10,12$.\label{fig:gammaCurv}}
\end{figure}
In the case of the spherical-wall the curvature dependence of the
surface tension is plotted in Fig. \ref{fig:gammaCurv}. There, results
for the 12-6 system at different temperatures are shown. At $T\apprge9$
fluid-wall surface tension is negative even at $R\rightarrow\infty$.
This negative sign of $\gamma$ is characteristic of systems with
repulsive interaction such as the hard sphere fluid. Fig. \ref{fig:gammaCurv}
is also related with the excess surface adsorption $\Gamma=\left(<n>-<n>_{\textrm{b}}\right)/A$.
Series expansion of $\Gamma$ up to order $z^{2}$ and $\rho_{\textrm{b}}^{2}$
are: $\Gamma=z^{2}\Delta\tau_{2}/A$$=\rho_{\textrm{b}}^{2}\Delta\tau_{2}/A$,
respectively. Thus, up to the order of Eq. (\ref{eq:gammaps}) it
is $\Gamma=-2\gamma/T$. This shows that isotherms of $\gamma(R)$
shown in Fig. \ref{fig:gammaCurv} also plot isotherms of $-\Gamma T/2$.
Naturally, the same apply to the planar case shown in Fig.\ref{fig:gammaPlanar}
and to the cylindrical one (not shown).

It must be noted that $\gamma(R)$ and $\Gamma$ depend on the adopted
surface of tension that we fixed at $r=R$ where $\rho(r)$ drops
to zero. This fixes the adopted reference region characterized by
measures $V$, $A$ and $R$. The effect of introducing a different
reference region on $\gamma(R)$ was systematically studied in Refs.
\cite{Urrutia_2014,Urrutia_2015} for the hard-sphere fluid and the
same approach applies to the LJ fluid. 
\begin{figure}
\centering{}\includegraphics[width=0.85\columnwidth]{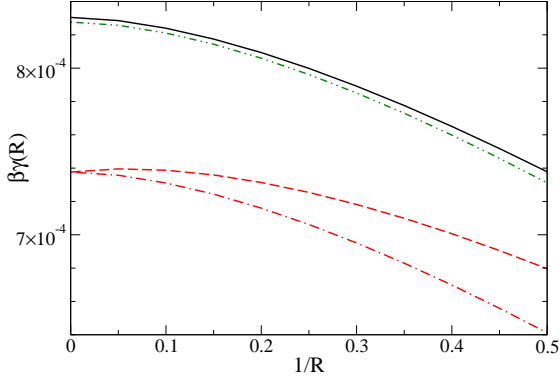}\caption{Curvature dependent surface tension scaled with the inverse temperature.
Continuous line shows our result {[}Eq. (\ref{eq:Gamrhob}){]}, dashed
line was extracted from Ref. \cite{Stewart_2005_b}. Other curves
are described in the text.\label{fig:betagammaR}}
\end{figure}

Stewart and Evans studied the interfacial properties of a hard spherical
cavity immersed in a fluid using effective interfacial potentials
and density functional theory. They used an interaction potential
between particles that contains both a hard sphere repulsion and an
attractive $-r^{-6}$ component, the latter similar to that appearing
in the 12-6 LJ potential.\cite{Stewart_2005_b} In Fig. 3 therein
it was presented a plot of $\beta\gamma$ as a function of $R^{-1}$
at $T=0.991$ and $\rho=0.018$. For comparison we present in Fig.
\ref{fig:betagammaR} the curve found by Stewart and Evans (dashed)
and our results for the 12-6 LJ gas obtained using Eq. (\ref{eq:Gamrhob})
at the same temperature and density (continuous). We observe an overall
discrepancy of $\sim10\%$ which is acceptable by virtue of the disparity
in the interaction model. Two major differences between both curves
account most of the observed discrepancy. On the one hand, the ordinate
at the origin i.e. the value of surface tension in the limit of planar
wall. On the other hand, the slope of curves at $R^{-1}\rightarrow0$
which is not zero for dashed curve. The difference in the observed
planar-wall surface tension is a direct consequence of the disparity
in the interaction model. Even though, the difference in the slope
is produced by our second order truncation that forces a zero slope.
In dot-dashed we present a version of dashed line compensated to give
$\delta(R^{-1}\rightarrow0)=0$. This last line was shifted an arbitrary
value and plotted in dot-dot-dashed to make clear the coincidence
with the obtained virial series result. In this case the shape is
identical which suggest that $\ln R/R^{2}$ and $R^{-2}$ terms are
not susceptible to the disparity of potentials.
\begin{figure}
\centering{}\includegraphics[width=0.85\columnwidth]{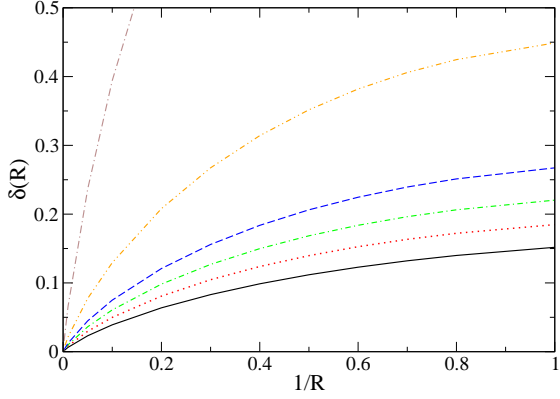}\caption{Radius dependent Tolman length of the 12-6 LJ fluid in contact with
a spherical wall (for both concave and convex shapes) and at various
temperatures. From bottom to top $T=1,2,3,4,6,8$.\label{fig:DeltaTol}}
\end{figure}

The fluid-substrate radius dependent Tolman Length, defined by $\delta(R)=\frac{R}{2}\left[1-\frac{\gamma(R)}{\gamma}\right]$,
measures the dependence of the surface tension with the curvature.
For all temperatures we found a positive $\delta(R)$, that goes to
zero at the planar limit and increases monotonously with $R^{-1}$.
Curve for $T=8$ increases monotonously until reaches the value $\delta(1)=1.33$.
Note that at Boile temperature $T=9.016$ where $\gamma$ goes to
zero $\delta(R)$ diverges.

\subsection{Curvature expansion\label{sub:CurvatureExp}}

We follow Ref. \cite{Blokhuis_2013c} to analyze the curvature expansion
of the surface tension. There, the analysis was done for the vapor
fluid interface. Helfrich\cite{Helfrich_1973} gives an expansion
of the surface tension in the curvature up to second order. Applied
to the sphere and cylinder symmetry Helfrich expansion of $\gamma(R)$
gives
\begin{eqnarray}
\gamma_{\textrm{s}}(R) & = & \gamma-\frac{2\delta\gamma}{R}+\frac{2\boldsymbol{k}+\bar{\boldsymbol{k}}}{R^{2}}+\ldots\:,\label{eq:gammaks}\\
\gamma_{\textrm{c}}(R) & = & \gamma-\frac{\delta\gamma}{R}+\frac{\boldsymbol{k}}{2R^{2}}+\ldots\:,\label{eq:gammakc}
\end{eqnarray}
where dots represent terms of $\mathcal{O}(R^{-3})$. The radius independent
Tolman length is $\delta(R\rightarrow\infty)=\delta$. Next term beyond
$\delta$ is related with the bending rigidity $\boldsymbol{k}$ related
with the square of the total curvature and the Gaussian rigidity $\bar{\boldsymbol{k}}$
associated with Gaussian curvature. On the basis of our results expansion
on Eqs. (\ref{eq:gammaks}, \ref{eq:gammakc}) are adequate for $k>6$.
Therefore, for $k>6$ we found $\delta=\mathcal{O}(\rho_{\textrm{b}}^{3})$,
\begin{eqnarray}
\boldsymbol{k} & = & -\frac{\pi}{32k}C_{6/k}(0)\,T\,\rho_{\textrm{b}}^{2}+\ldots\:,\label{eq:kkbar}\\
\bar{\boldsymbol{k}} & = & \frac{\pi}{48}C_{6/k}(0)\,T\,\rho_{\textrm{b}}^{2}+\ldots\:,
\end{eqnarray}
where dots represent terms of order $\mathcal{O}(\rho_{\textrm{b}}^{3})$.
Note that even at order $\rho_{\textrm{b}}^{2}$ both rigid constants
have a non trivial dependence on $T$. It is interesting to calculate
the quotient between $\boldsymbol{k}$ and $\bar{\boldsymbol{k}}$
which gives for all $k>6$ 
\begin{equation}
\boldsymbol{k}/\bar{\boldsymbol{k}}=-3/2\:.\label{eq:qkk}
\end{equation}
Remarkably, it is a universal value in the sense that it is independent
of both $k$ and the state variables $T$ and $\rho_{\textrm{b}}$.
It is trivial to verify that this relation also applies to HS and
SW particles.\cite{Urrutia_2014}

For long-ranged interactions as in the case of 12-6 LJ the existence
of the discussed logarithmic terms makes the Helfrich expansion\cite{Helfrich_1973}
of $\gamma(R)$ in power of $R^{-1}$ no longer valid. Thus, for $k=6$
instead of the Helfrich-based expression Eqs. (\ref{eq:gammaks},
\ref{eq:gammakc}), one obtains for the spherical and cylindrical
walls 
\begin{eqnarray}
\gamma_{\textrm{s}}(R) & = & \gamma-\frac{2\delta\gamma}{R}+\left(2\boldsymbol{k}+\bar{\boldsymbol{k}}\right)\frac{\ln R^{-1}}{R^{2}}+\ldots\:,\label{eq:gammaksLR}\\
\gamma_{\textrm{c}}(R) & = & \gamma-\frac{\delta\gamma}{R}+\boldsymbol{k}\frac{\ln R^{-1}}{2R^{2}}+\ldots\:.\label{eq:gammakcLR}
\end{eqnarray}
Here dots represent terms of $\mathcal{O}(R^{-2})$. Again, bending
and Gaussian rigidities were identified with the next order terms
beyond $\delta$. We obtain for the series expansion beyond the null
Tolman length 
\begin{equation}
\boldsymbol{k}=\frac{\pi}{8}\rho_{\textrm{b}}^{2}+\ldots\:\:\:\textrm{ and }\:\:\:\bar{\boldsymbol{k}}=-\frac{\pi}{12}\rho_{\textrm{b}}^{2}+\ldots\:\:,\label{eq:kkbarLR}
\end{equation}
where dots represent terms of order $\mathcal{O}(\rho_{\textrm{b}}^{3})$.
In this case both rigidities are temperature independent. Even for
$k=6$ the quotient gives the same result $\boldsymbol{k}/\bar{\boldsymbol{k}}=-3/2$,
found in Eq. (\ref{eq:qkk}). In fact, the origin of this fundamental
value is purely \emph{geometrical} and lies in Eq. (\ref{eq:x2c2}).
Expressions identical to those given in Eq. (\ref{eq:kkbarLR}), but
with the difference between bulk densities in liquid and vapor phases
instead of $\rho_{\textrm{b}}$, were found previously for the free
liquid-vapor interface of the full (uncut) 12-6 LJ fluid.\cite{Blokhuis_2013c}

Thus, essentially any pair interacting potential, including all the
finite range potentials (e.g. the cut and shifted 12-6 LJ, HS, SW
and square-shoulder, potentials) produce the same value for the ratio
$\boldsymbol{k}/\bar{\boldsymbol{k}}$ at low density. We also calculate
the quotient of the next to $R^{-1}$ terms between spherical and
cylindrical cases, the ratio $\left(2\boldsymbol{k}+\bar{\boldsymbol{k}}\right)/\left(\boldsymbol{k}/2\right)=-8/3$.
Again it has a purely geometrical origin and has the advantage of
being independent of the assumptions of a Helfrich-based expression
for $\gamma(R)$. This result is in line with that found numerically
using a (second-virial approximation) DFT\cite{Reindl_2015} for all
the studied potentials: LJ, SW, square-shoulder and Yukawa, all of
them cut at finite range. The same \emph{geometrical} status claimed
for $\boldsymbol{k}/\bar{\boldsymbol{k}}$ corresponds to the result
$\delta=0+\mathcal{O}(\rho^{3})$ that is directly derivable from
Eq. (\ref{eq:dTau2}) and applies to essentially any pair potential.

Based on the Hadwiger theorem it was proposed that bending constant
could be zero,\cite{Konig_2005} and thus, is unnecessary to include
it in the expansion of $\gamma(R)$. Eq. (\ref{eq:qkk}) shows that
the inaccuracy introduced by truncation of the bending rigidity term
in Eqs. (\ref{eq:gammaks}, \ref{eq:gammakc}) is of the same order
than the Gaussian rigidity term (at least for hard walls), and therefore
is not well justified from the numerical standpoint.

\section{Conclusions\label{sec:Summary}}

We give a simple and concise presentation of statistical mechanics
for inhomogeneous fluid systems that is appropriate to discuss virial
series in powers of the activity. The advantage of the adopted framework
is highlighted by showing short and explicit expressions of virial
series for the free energy and one- and two-body distribution functions.
Our approach, that avoids the introduction of \emph{a priori} assumptions
about the free energy is, in fact, a selection of different formulations
and ideas taken from Bellemans, Sokolowski and Stecki, and Rowlinson,
that we managed to assimilate and develop a synthetical representation.

Our point of view heightens the relevance of cluster integrals by
their generalization to inhomogeneous fluid-type systems. These cluster
integrals, that reduce to the Mayer ones in the case of homogeneous
fluids, are the coefficients of the series expansion in the activity
of the free energy. Extensions of the concept of cluster integral
and cluster integrand enable us to analyze under the same approach
the residual terms of distribution functions. Virial series in power
of either the bulk and mean densities (the bulk density as in Bellemans
approach, or the mean density of the system as adopted by Rowlinson)
are thus considered as two of many possible choices for the independent
variable in the power series representation of free energy. It should
noted that expansions in the activity have shown to be simpler to
analyze when different conventions for the reference region (its volume,
area and shape) are utilized.\cite{Urrutia_2014,Urrutia_2015}

Second order terms i.e. the second cluster integral and the second
order residue of one particle distribution were analyzed in detail
when the system is confined by hard walls of an arbitrary shape. To
do so we incorporated the advances developed by Bellemans, Sokolowski
and Stecki. By limiting the study cases of the applied external potential
and the order of the expansion we were able to shift the load to the
solution of cluster integrals. This is not minor since it reduces
an originally general and quite hard to address framework to a more
straightforward method. It should not be overlooked that the hypothesis
leading to Eq. (\ref{eq:Tau2A2}) for $\Delta\tau_{2}$ are common
to a lot of systems of major interest, at least in an approximate
manner. Then, $\Delta\tau_{2}$ equation becomes a rather powerful
tool for tackling inhomogeneous systems in a wide spectrum, specially
for direct numerical solving.

By analytically solving the simpler situations for $\Delta\tau_{2}$
we were able to expose the volume, area and other terms that contribute,
acquiring the capacity to discriminate between bulk terms and a hierarchy
of inhomogeneous terms that characterizes the curvature dependence.
Particularly, we focused in confining regions with a constant curvature
boundary: planar, spherical and cylindrical cases. As a simple application
of our findings to a non-trivial problem, we analyzed the second cluster
integral for the confined LJ system. We evaluated analytically the
temperature and radius dependence of $\tau_{2}$ and $\left\llbracket \rho^{(2)}(\mathbf{r})\right\rrbracket $.
The 12-6 LJ system was considered but also the more general 2$k$-$k$
LJ potential. It was found that second cluster integral of the 12-6
LJ system contains a bulk, a surface, and also a non-analytic dependence
with $R$. The latter are, in a spherical confinement a $\ln R$ term
and, in cylindrical confinement a $L\ln R/R$ term. For $k\geq7$
these logarithmic dependencies are absent (sphere) or may appear at
higher order in $R^{-1}$ (cylinder) but in all cases a series of
terms proportional to negative powers of $R$ were also obtained.
We obtained the free energy of the inhomogeneous systems by truncation
of the virial series order at order $z^{2}$ and $\rho_{\textrm{b}}^{2}$,
that directly maps our findings on $\tau_{2}$ to free energy. The
existence of Log terms in the free energy of fluids in contact with
hard spherical surfaces was hypothesized by Henderson and later discussed
by Stecki and col. \cite{Henderson_1983,Stecki_1990,Poniewierski_1997,Samborski_1993}.
Our results demonstrate this conjecture for the 12-6 LJ system.

The fluid-substrate surface tension was also analyzed using second
order truncated virial series. In the planar case we found an exact
expression that describes $\gamma(T)$ for all $k\geq6$. We evaluated
the temperature below that the surface tension becomes negative. For
$k=6$ it is $6.8\times T_{c}$. The prefactor decreases with $k$
being $2.2\times T_{c}$ for $k=18$ that corresponds to short range
potentials proper of colloidal particles. Based on the virial series
approach the leading order curvature correction to the surface tension
$\gamma(R)$ for all the $2k$-$k$ LJ fluids in contact with spherical
and cylindrical surfaces was found analytically, at order two in density.
This correction is the same when the system is inside of the surface
or outside of it. For $k=6$ in the case of both spherical and cylindrical
confinement surface tension scales with $\ln R/R^{2}$. For the $k\geq7$
the first correction is order $R^{-2}$. In all cases the correction
is negative for high temperatures. The truncation of the 12-6 LJ potential
produces a significant change in the dependence of $\gamma(R)$ with
$R$, vanishing the $\ln R/R^{2}$ dependence and producing a simple
correction. We observed that the term of order $R^{-1}$ is zero for
$k\geq6$. This shows that Tolman length is zero up to order $\rho_{\textrm{b}}^{2}$
but it should appear at higher order ones, probably at order $\rho_{\textrm{b}}^{3}$.

The curvature dependence of surface tension for fluid interfaces is
a highly studied issue.\cite{Blokhuis_2013c,Corti_2011,HansenGoos_2014,Troster_2012,Baidakov_2009,Wilhelmsen_2015,vanGiessen_2009,Blokhuis_2013,Urrutia_2014,Baidakov_2014b}
In particular, the amplitude and sign of bending and Gaussian rigidity
constants are a matter of discussion. We evaluated analytically both
rigidities, $\boldsymbol{k}$ and $\bar{\boldsymbol{k}}$, at order
$z^{2}$ and $\rho_{\textrm{b}}^{2}$. For $k=6$ both bending and
Gaussian rigidities are independent of temperature, being $\boldsymbol{k}>0$
and $\bar{\boldsymbol{k}}$<0. Other values of $k$ are characterized
by a temperature where both rigidities change its sign. For all $k\geq6$
we obtained the universal ratio $\boldsymbol{k}/\bar{\boldsymbol{k}}=-3/2$
which is a thermodynamic result based on pure geometrical grounds.
This value is exact when terms of order $\rho_{\textrm{b}}^{3}$ are
truncated from the EOS of the system. We obtain the same result for
any finite range potential.

Previous works have discussed the existence of $\ln R$ non-analytic
dependence of the surface tension with the curvature when dispersion
forces are present and multiple techniques were used with this purpose
including DFT, MonteCarlo, Molecular dynamics and effective Hamiltonian.\cite{Schmitz_2014,Schmitz_2014b}
These terms were found at the gas-liquid interface of droplets and
bubbles,\cite{Blokhuis_2013c} and at the curved wall-fluid interface.\cite{Blokhuis_2007,Stewart_2005_b}
In wetting and drying at curved surfaces it was also identified.\cite{Evans_2003,Evans_2004,Stewart_2005,Stewart_2005_b}
In all those cases the magnitude of this term is indirectly evaluated:
it may involve the truncation of the interaction potential, the fitting
of density profiles and/or surface tension curves, the use of approximate
EOS for the bulk system or more than one of this approximations. This
yields results that require deeper testing. Our analytical approach
is a contribution in that direction. 
\begin{acknowledgments}
This work was supported by Argentina Grants ANPCyT PICT-2011-1887,
and CONICET PIP-112-200801-00403.
\end{acknowledgments}


\begin{thebibliography}{93}%
\makeatletter
\providecommand \@ifxundefined [1]{%
 \@ifx{#1\undefined}
}%
\providecommand \@ifnum [1]{%
 \ifnum #1\expandafter \@firstoftwo
 \else \expandafter \@secondoftwo
 \fi
}%
\providecommand \@ifx [1]{%
 \ifx #1\expandafter \@firstoftwo
 \else \expandafter \@secondoftwo
 \fi
}%
\providecommand \natexlab [1]{#1}%
\providecommand \enquote  [1]{``#1''}%
\providecommand \bibnamefont  [1]{#1}%
\providecommand \bibfnamefont [1]{#1}%
\providecommand \citenamefont [1]{#1}%
\providecommand \href@noop [0]{\@secondoftwo}%
\providecommand \href [0]{\begingroup \@sanitize@url \@href}%
\providecommand \@href[1]{\@@startlink{#1}\@@href}%
\providecommand \@@href[1]{\endgroup#1\@@endlink}%
\providecommand \@sanitize@url [0]{\catcode `\\12\catcode `\$12\catcode
  `\&12\catcode `\#12\catcode `\^12\catcode `\_12\catcode `\%12\relax}%
\providecommand \@@startlink[1]{}%
\providecommand \@@endlink[0]{}%
\providecommand \url  [0]{\begingroup\@sanitize@url \@url }%
\providecommand \@url [1]{\endgroup\@href {#1}{\urlprefix }}%
\providecommand \urlprefix  [0]{URL }%
\providecommand \Eprint [0]{\href }%
\providecommand \doibase [0]{http://dx.doi.org/}%
\providecommand \selectlanguage [0]{\@gobble}%
\providecommand \bibinfo  [0]{\@secondoftwo}%
\providecommand \bibfield  [0]{\@secondoftwo}%
\providecommand \translation [1]{[#1]}%
\providecommand \BibitemOpen [0]{}%
\providecommand \bibitemStop [0]{}%
\providecommand \bibitemNoStop [0]{.\EOS\space}%
\providecommand \EOS [0]{\spacefactor3000\relax}%
\providecommand \BibitemShut  [1]{\csname bibitem#1\endcsname}%
\let\auto@bib@innerbib\@empty
\bibitem [{\citenamefont {Waals}(1873)}]{vanderWaals_1873}%
  \BibitemOpen
  \bibfield  {author} {\bibinfo {author} {\bibfnamefont {J.~D. v.~d.}\
  \bibnamefont {Waals}},\ }\href@noop {} {\bibfield  {journal} {\bibinfo
  {journal} {Dissertation, Leiden}\ } (\bibinfo {year} {1873})}\BibitemShut
  {NoStop}%
\bibitem [{\citenamefont {Laar}(1899)}]{vanLaar_1899}%
  \BibitemOpen
  \bibfield  {author} {\bibinfo {author} {\bibfnamefont {J.~J.~v.}\
  \bibnamefont {Laar}},\ }\href@noop {} {\bibfield  {journal} {\bibinfo
  {journal} {Amsterdam Akad. Versl.}\ }\textbf {\bibinfo {volume} {7}},\
  \bibinfo {pages} {350} (\bibinfo {year} {1899})}\BibitemShut {NoStop}%
\bibitem [{\citenamefont {Ushcats}(2013)}]{Ushcats_2013}%
  \BibitemOpen
  \bibfield  {author} {\bibinfo {author} {\bibfnamefont {M.~V.}\ \bibnamefont
  {Ushcats}},\ }\href {\doibase 10.1063/1.4793407} {\bibfield  {journal}
  {\bibinfo  {journal} {The Journal of Chemical Physics}\ }\textbf {\bibinfo
  {volume} {138}},\ \bibinfo {eid} {094309} (\bibinfo {year}
  {2013})}\BibitemShut {NoStop}%
\bibitem [{\citenamefont {Hellmann}\ and\ \citenamefont
  {Bich}(2011)}]{Hellmann_2011}%
  \BibitemOpen
  \bibfield  {author} {\bibinfo {author} {\bibfnamefont {R.}~\bibnamefont
  {Hellmann}}\ and\ \bibinfo {author} {\bibfnamefont {E.}~\bibnamefont
  {Bich}},\ }\href {\doibase 10.1063/1.3626524} {\bibfield  {journal} {\bibinfo
   {journal} {Journal of Chemical Physics}\ }\textbf {\bibinfo {volume}
  {135}},\ \bibinfo {pages} {084117} (\bibinfo {year} {2011})}\BibitemShut
  {NoStop}%
\bibitem [{\citenamefont {Shaul}\ \emph {et~al.}(2012)\citenamefont {Shaul},
  \citenamefont {Schultz},\ and\ \citenamefont {Kofke}}]{Shaul_2012}%
  \BibitemOpen
  \bibfield  {author} {\bibinfo {author} {\bibfnamefont {K.~R.~S.}\
  \bibnamefont {Shaul}}, \bibinfo {author} {\bibfnamefont {A.~J.}\ \bibnamefont
  {Schultz}}, \ and\ \bibinfo {author} {\bibfnamefont {D.~A.}\ \bibnamefont
  {Kofke}},\ }\href {\doibase 10.1063/1.4764857} {\bibfield  {journal}
  {\bibinfo  {journal} {The Journal of Chemical Physics}\ }\textbf {\bibinfo
  {volume} {137}},\ \bibinfo {eid} {184101} (\bibinfo {year}
  {2012})}\BibitemShut {NoStop}%
\bibitem [{\citenamefont {Gazzillo}\ and\ \citenamefont
  {Pini}(2013)}]{Gazzillo_2013}%
  \BibitemOpen
  \bibfield  {author} {\bibinfo {author} {\bibfnamefont {D.}~\bibnamefont
  {Gazzillo}}\ and\ \bibinfo {author} {\bibfnamefont {D.}~\bibnamefont
  {Pini}},\ }\href {\doibase 10.1063/1.4825174} {\bibfield  {journal} {\bibinfo
   {journal} {The Journal of Chemical Physics}\ }\textbf {\bibinfo {volume}
  {139}},\ \bibinfo {eid} {164501} (\bibinfo {year} {2013})}\BibitemShut
  {NoStop}%
\bibitem [{\citenamefont {Beltran-Heredia}\ and\ \citenamefont
  {Santos}(2014)}]{BeltranHeredia_2014}%
  \BibitemOpen
  \bibfield  {author} {\bibinfo {author} {\bibfnamefont {E.}~\bibnamefont
  {Beltran-Heredia}}\ and\ \bibinfo {author} {\bibfnamefont {A.}~\bibnamefont
  {Santos}},\ }\href {\doibase 10.1063/1.4870011} {\bibfield  {journal}
  {\bibinfo  {journal} {The Journal of Chemical Physics}\ }\textbf {\bibinfo
  {volume} {140}},\ \bibinfo {eid} {134507} (\bibinfo {year}
  {2014})}\BibitemShut {NoStop}%
\bibitem [{\citenamefont {Korden}(2012)}]{Korden_2012}%
  \BibitemOpen
  \bibfield  {author} {\bibinfo {author} {\bibfnamefont {S.}~\bibnamefont
  {Korden}},\ }\href {\doibase 10.1103/PhysRevE.85.041150} {\bibfield
  {journal} {\bibinfo  {journal} {Phys. Rev. E}\ }\textbf {\bibinfo {volume}
  {85}},\ \bibinfo {pages} {041150} (\bibinfo {year} {2012})}\BibitemShut
  {NoStop}%
\bibitem [{\citenamefont {Dudowicz}\ \emph {et~al.}(2015)\citenamefont
  {Dudowicz}, \citenamefont {Freed},\ and\ \citenamefont
  {Douglas}}]{Dudowicz_2015}%
  \BibitemOpen
  \bibfield  {author} {\bibinfo {author} {\bibfnamefont {J.}~\bibnamefont
  {Dudowicz}}, \bibinfo {author} {\bibfnamefont {K.~F.}\ \bibnamefont {Freed}},
  \ and\ \bibinfo {author} {\bibfnamefont {J.~F.}\ \bibnamefont {Douglas}},\
  }\href {\doibase 10.1063/1.4935705} {\bibfield  {journal} {\bibinfo
  {journal} {The Journal of Chemical Physics}\ }\textbf {\bibinfo {volume}
  {143}},\ \bibinfo {eid} {194901} (\bibinfo {year} {2015})}\BibitemShut
  {NoStop}%
\bibitem [{\citenamefont {Koga}\ \emph {et~al.}(2015)\citenamefont {Koga},
  \citenamefont {Holten},\ and\ \citenamefont {Widom}}]{Koga_2015}%
  \BibitemOpen
  \bibfield  {author} {\bibinfo {author} {\bibfnamefont {K.}~\bibnamefont
  {Koga}}, \bibinfo {author} {\bibfnamefont {V.}~\bibnamefont {Holten}}, \ and\
  \bibinfo {author} {\bibfnamefont {B.}~\bibnamefont {Widom}},\ }\href
  {\doibase 10.1021/acs.jpcb.5b07685} {\bibfield  {journal} {\bibinfo
  {journal} {The Journal of Physical Chemistry B}\ }\textbf {\bibinfo {volume}
  {119}},\ \bibinfo {pages} {13391} (\bibinfo {year} {2015})}\BibitemShut
  {NoStop}%
\bibitem [{\citenamefont {L\'opez~de Haro}\ \emph {et~al.}(2015)\citenamefont
  {L\'opez~de Haro}, \citenamefont {Tejero}, \citenamefont {Santos},
  \citenamefont {Yuste}, \citenamefont {Fiumara},\ and\ \citenamefont
  {Saija}}]{LopezdeHaro_2015}%
  \BibitemOpen
  \bibfield  {author} {\bibinfo {author} {\bibfnamefont {M.}~\bibnamefont
  {L\'opez~de Haro}}, \bibinfo {author} {\bibfnamefont {C.~F.}\ \bibnamefont
  {Tejero}}, \bibinfo {author} {\bibfnamefont {A.}~\bibnamefont {Santos}},
  \bibinfo {author} {\bibfnamefont {S.~B.}\ \bibnamefont {Yuste}}, \bibinfo
  {author} {\bibfnamefont {G.}~\bibnamefont {Fiumara}}, \ and\ \bibinfo
  {author} {\bibfnamefont {F.}~\bibnamefont {Saija}},\ }\href {\doibase
  10.1063/1.4904891} {\bibfield  {journal} {\bibinfo  {journal} {The Journal of
  Chemical Physics}\ }\textbf {\bibinfo {volume} {142}},\ \bibinfo {eid}
  {014902} (\bibinfo {year} {2015})}\BibitemShut {NoStop}%
\bibitem [{\citenamefont {Hutem}\ and\ \citenamefont
  {Boonchui}(2012)}]{Hutem_2012}%
  \BibitemOpen
  \bibfield  {author} {\bibinfo {author} {\bibfnamefont {A.}~\bibnamefont
  {Hutem}}\ and\ \bibinfo {author} {\bibfnamefont {S.}~\bibnamefont
  {Boonchui}},\ }\href {\doibase 10.1007/s10910-011-9966-5} {\bibfield
  {journal} {\bibinfo  {journal} {Journal of Mathematical Chemistry}\ }\textbf
  {\bibinfo {volume} {50}},\ \bibinfo {pages} {1262} (\bibinfo {year}
  {2012})}\BibitemShut {NoStop}%
\bibitem [{\citenamefont {Schultz}\ and\ \citenamefont
  {Kofke}(2010)}]{Schultz_2010}%
  \BibitemOpen
  \bibfield  {author} {\bibinfo {author} {\bibfnamefont {A.~J.}\ \bibnamefont
  {Schultz}}\ and\ \bibinfo {author} {\bibfnamefont {D.~A.}\ \bibnamefont
  {Kofke}},\ }\href {\doibase 10.1063/1.3486085} {\bibfield  {journal}
  {\bibinfo  {journal} {The Journal of Chemical Physics}\ }\textbf {\bibinfo
  {volume} {133}},\ \bibinfo {eid} {104101} (\bibinfo {year}
  {2010})}\BibitemShut {NoStop}%
\bibitem [{\citenamefont {Ashbaugh}\ \emph {et~al.}(2015)\citenamefont
  {Ashbaugh}, \citenamefont {Weiss}, \citenamefont {Williams}, \citenamefont
  {Meng},\ and\ \citenamefont {Surampudi}}]{Ashbaugh_2015}%
  \BibitemOpen
  \bibfield  {author} {\bibinfo {author} {\bibfnamefont {H.~S.}\ \bibnamefont
  {Ashbaugh}}, \bibinfo {author} {\bibfnamefont {K.}~\bibnamefont {Weiss}},
  \bibinfo {author} {\bibfnamefont {S.~M.}\ \bibnamefont {Williams}}, \bibinfo
  {author} {\bibfnamefont {B.}~\bibnamefont {Meng}}, \ and\ \bibinfo {author}
  {\bibfnamefont {L.~N.}\ \bibnamefont {Surampudi}},\ }\href {\doibase
  10.1021/acs.jpcb.5b02056} {\bibfield  {journal} {\bibinfo  {journal} {The
  Journal of Physical Chemistry B}\ }\textbf {\bibinfo {volume} {119}},\
  \bibinfo {pages} {6280} (\bibinfo {year} {2015})}\BibitemShut {NoStop}%
\bibitem [{\citenamefont {Virga}(2013)}]{Virga_2013}%
  \BibitemOpen
  \bibfield  {author} {\bibinfo {author} {\bibfnamefont {E.~G.}\ \bibnamefont
  {Virga}},\ }\href {http://stacks.iop.org/0953-8984/25/i=46/a=465109}
  {\bibfield  {journal} {\bibinfo  {journal} {Journal of Physics: Condensed
  Matter}\ }\textbf {\bibinfo {volume} {25}},\ \bibinfo {pages} {465109}
  (\bibinfo {year} {2013})}\BibitemShut {NoStop}%
\bibitem [{\citenamefont {Henderson}(2011)}]{HendersonD_2011}%
  \BibitemOpen
  \bibfield  {author} {\bibinfo {author} {\bibfnamefont {D.}~\bibnamefont
  {Henderson}},\ }\href {\doibase 10.1063/1.3615723} {\bibfield  {journal}
  {\bibinfo  {journal} {Journal of Chemical Physics}\ }\textbf {\bibinfo
  {volume} {135}},\ \bibinfo {pages} {044514} (\bibinfo {year}
  {2011})}\BibitemShut {NoStop}%
\bibitem [{\citenamefont {Philipse}\ and\ \citenamefont
  {Kuipers}(2010)}]{Philipse_2010}%
  \BibitemOpen
  \bibfield  {author} {\bibinfo {author} {\bibfnamefont {A.~P.}\ \bibnamefont
  {Philipse}}\ and\ \bibinfo {author} {\bibfnamefont {B.~W.~M.}\ \bibnamefont
  {Kuipers}},\ }\href {http://stacks.iop.org/0953-8984/22/i=32/a=325104}
  {\bibfield  {journal} {\bibinfo  {journal} {Journal of Physics: Condensed
  Matter}\ }\textbf {\bibinfo {volume} {22}},\ \bibinfo {pages} {325104}
  (\bibinfo {year} {2010})}\BibitemShut {NoStop}%
\bibitem [{\citenamefont {Mamedov}\ and\ \citenamefont
  {Somuncu}(2015)}]{Mamedov_2015}%
  \BibitemOpen
  \bibfield  {author} {\bibinfo {author} {\bibfnamefont {B.}~\bibnamefont
  {Mamedov}}\ and\ \bibinfo {author} {\bibfnamefont {E.}~\bibnamefont
  {Somuncu}},\ }\href {\doibase 10.1016/j.physa.2014.11.014} {\bibfield
  {journal} {\bibinfo  {journal} {Physica A: Statistical Mechanics and its
  Applications}\ }\textbf {\bibinfo {volume} {420}},\ \bibinfo {pages} {246 }
  (\bibinfo {year} {2015})}\BibitemShut {NoStop}%
\bibitem [{\citenamefont {Santos}\ \emph {et~al.}(2015)\citenamefont {Santos},
  \citenamefont {L\'opez~de Haro}, \citenamefont {Fiumara},\ and\ \citenamefont
  {Saija}}]{Santos_2015}%
  \BibitemOpen
  \bibfield  {author} {\bibinfo {author} {\bibfnamefont {A.}~\bibnamefont
  {Santos}}, \bibinfo {author} {\bibfnamefont {M.}~\bibnamefont {L\'opez~de
  Haro}}, \bibinfo {author} {\bibfnamefont {G.}~\bibnamefont {Fiumara}}, \ and\
  \bibinfo {author} {\bibfnamefont {F.}~\bibnamefont {Saija}},\ }\href
  {\doibase 10.1063/1.4922031} {\bibfield  {journal} {\bibinfo  {journal} {The
  Journal of Chemical Physics}\ }\textbf {\bibinfo {volume} {142}},\ \bibinfo
  {eid} {224903} (\bibinfo {year} {2015})}\BibitemShut {NoStop}%
\bibitem [{\citenamefont {Baidakov}\ \emph {et~al.}(2012)\citenamefont
  {Baidakov}, \citenamefont {Protsenko},\ and\ \citenamefont
  {Kozlova}}]{Baidakov_2012}%
  \BibitemOpen
  \bibfield  {author} {\bibinfo {author} {\bibfnamefont {V.~G.}\ \bibnamefont
  {Baidakov}}, \bibinfo {author} {\bibfnamefont {S.~P.}\ \bibnamefont
  {Protsenko}}, \ and\ \bibinfo {author} {\bibfnamefont {Z.~R.}\ \bibnamefont
  {Kozlova}},\ }\href {\doibase 10.1063/1.4758806} {\bibfield  {journal}
  {\bibinfo  {journal} {The Journal of Chemical Physics}\ }\textbf {\bibinfo
  {volume} {137}},\ \bibinfo {eid} {164507} (\bibinfo {year}
  {2012})}\BibitemShut {NoStop}%
\bibitem [{\citenamefont {Baidakov}\ and\ \citenamefont
  {Protsenko}(2014{\natexlab{a}})}]{Baidakov_2014c}%
  \BibitemOpen
  \bibfield  {author} {\bibinfo {author} {\bibfnamefont {V.~G.}\ \bibnamefont
  {Baidakov}}\ and\ \bibinfo {author} {\bibfnamefont {S.~P.}\ \bibnamefont
  {Protsenko}},\ }\href {\doibase 10.1063/1.4895624} {\bibfield  {journal}
  {\bibinfo  {journal} {The Journal of Chemical Physics}\ }\textbf {\bibinfo
  {volume} {141}},\ \bibinfo {eid} {114503} (\bibinfo {year}
  {2014}{\natexlab{a}})}\BibitemShut {NoStop}%
\bibitem [{\citenamefont {Baidakov}\ and\ \citenamefont
  {Protsenko}(2014{\natexlab{b}})}]{Baidakov_2014}%
  \BibitemOpen
  \bibfield  {author} {\bibinfo {author} {\bibfnamefont {V.~G.}\ \bibnamefont
  {Baidakov}}\ and\ \bibinfo {author} {\bibfnamefont {S.~P.}\ \bibnamefont
  {Protsenko}},\ }\href {\doibase 10.1063/1.4880958} {\bibfield  {journal}
  {\bibinfo  {journal} {The Journal of Chemical Physics}\ }\textbf {\bibinfo
  {volume} {140}},\ \bibinfo {eid} {214506} (\bibinfo {year}
  {2014}{\natexlab{b}})}\BibitemShut {NoStop}%
\bibitem [{\citenamefont {Baidakov}\ and\ \citenamefont
  {Bobrov}(2014)}]{Baidakov_2014b}%
  \BibitemOpen
  \bibfield  {author} {\bibinfo {author} {\bibfnamefont {V.~G.}\ \bibnamefont
  {Baidakov}}\ and\ \bibinfo {author} {\bibfnamefont {K.~S.}\ \bibnamefont
  {Bobrov}},\ }\href {\doibase 10.1063/1.4874644} {\bibfield  {journal}
  {\bibinfo  {journal} {The Journal of Chemical Physics}\ }\textbf {\bibinfo
  {volume} {140}},\ \bibinfo {eid} {184506} (\bibinfo {year}
  {2014})}\BibitemShut {NoStop}%
\bibitem [{\citenamefont {Heyes}\ and\ \citenamefont
  {Bra\'nka}(2015)}]{Heyes_2015}%
  \BibitemOpen
  \bibfield  {author} {\bibinfo {author} {\bibfnamefont {D.~M.}\ \bibnamefont
  {Heyes}}\ and\ \bibinfo {author} {\bibfnamefont {A.~C.}\ \bibnamefont
  {Bra\'nka}},\ }\href {\doibase 10.1063/1.4937487} {\bibfield  {journal}
  {\bibinfo  {journal} {The Journal of Chemical Physics}\ }\textbf {\bibinfo
  {volume} {143}},\ \bibinfo {eid} {234504} (\bibinfo {year}
  {2015})}\BibitemShut {NoStop}%
\bibitem [{\citenamefont {Blokhuis}\ and\ \citenamefont {van
  Giessen}(2013)}]{Blokhuis_2013c}%
  \BibitemOpen
  \bibfield  {author} {\bibinfo {author} {\bibfnamefont {E.~M.}\ \bibnamefont
  {Blokhuis}}\ and\ \bibinfo {author} {\bibfnamefont {A.~E.}\ \bibnamefont {van
  Giessen}},\ }\href {http://stacks.iop.org/0953-8984/25/i=22/a=225003}
  {\bibfield  {journal} {\bibinfo  {journal} {Journal of Physics: Condensed
  Matter}\ }\textbf {\bibinfo {volume} {25}},\ \bibinfo {pages} {225003}
  (\bibinfo {year} {2013})}\BibitemShut {NoStop}%
\bibitem [{\citenamefont {Blokhuis}\ and\ \citenamefont
  {Kuipers}(2007)}]{Blokhuis_2007}%
  \BibitemOpen
  \bibfield  {author} {\bibinfo {author} {\bibfnamefont {E.~M.}\ \bibnamefont
  {Blokhuis}}\ and\ \bibinfo {author} {\bibfnamefont {J.}~\bibnamefont
  {Kuipers}},\ }\href {\doibase 10.1063/1.2434161} {\bibfield  {journal}
  {\bibinfo  {journal} {The Journal of Chemical Physics}\ }\textbf {\bibinfo
  {volume} {126}},\ \bibinfo {eid} {054702} (\bibinfo {year}
  {2007})}\BibitemShut {NoStop}%
\bibitem [{\citenamefont {van Giessen}\ and\ \citenamefont
  {Blokhuis}(2002)}]{vanGiessen_2002}%
  \BibitemOpen
  \bibfield  {author} {\bibinfo {author} {\bibfnamefont {A.~E.}\ \bibnamefont
  {van Giessen}}\ and\ \bibinfo {author} {\bibfnamefont {E.~M.}\ \bibnamefont
  {Blokhuis}},\ }\href {\doibase 10.1063/1.1423617} {\bibfield  {journal}
  {\bibinfo  {journal} {The Journal of Chemical Physics}\ }\textbf {\bibinfo
  {volume} {116}},\ \bibinfo {pages} {302} (\bibinfo {year}
  {2002})}\BibitemShut {NoStop}%
\bibitem [{\citenamefont {Gibbons}\ and\ \citenamefont
  {Steele}(1971)}]{Gibbons_1971}%
  \BibitemOpen
  \bibfield  {author} {\bibinfo {author} {\bibfnamefont {R.}~\bibnamefont
  {Gibbons}}\ and\ \bibinfo {author} {\bibfnamefont {W.}~\bibnamefont
  {Steele}},\ }\href {\doibase 10.1080/00268977100101081} {\bibfield  {journal}
  {\bibinfo  {journal} {Molecular Physics}\ }\textbf {\bibinfo {volume} {20}},\
  \bibinfo {pages} {1099} (\bibinfo {year} {1971})}\BibitemShut {NoStop}%
\bibitem [{\citenamefont {Caligaris}\ and\ \citenamefont
  {Rodriguez}(1971)}]{Caligaris_1971}%
  \BibitemOpen
  \bibfield  {author} {\bibinfo {author} {\bibfnamefont {R.}~\bibnamefont
  {Caligaris}}\ and\ \bibinfo {author} {\bibfnamefont {A.}~\bibnamefont
  {Rodriguez}},\ }\href {\doibase 10.1080/00268977100103441} {\bibfield
  {journal} {\bibinfo  {journal} {Molecular Physics}\ }\textbf {\bibinfo
  {volume} {22}},\ \bibinfo {pages} {1131} (\bibinfo {year}
  {1971})}\BibitemShut {NoStop}%
\bibitem [{\citenamefont {Schultz}\ and\ \citenamefont
  {Kofke}(2009{\natexlab{a}})}]{Schultz_2009_b}%
  \BibitemOpen
  \bibfield  {author} {\bibinfo {author} {\bibfnamefont {A.~J.}\ \bibnamefont
  {Schultz}}\ and\ \bibinfo {author} {\bibfnamefont {D.~A.}\ \bibnamefont
  {Kofke}},\ }\href {\doibase 10.1080/00268970903267053} {\bibfield  {journal}
  {\bibinfo  {journal} {Molecular Physics}\ }\textbf {\bibinfo {volume}
  {107}},\ \bibinfo {pages} {2309} (\bibinfo {year}
  {2009}{\natexlab{a}})}\BibitemShut {NoStop}%
\bibitem [{\citenamefont {Feng}\ \emph {et~al.}(2015)\citenamefont {Feng},
  \citenamefont {Schultz}, \citenamefont {Chaudhary},\ and\ \citenamefont
  {Kofke}}]{Feng_2015}%
  \BibitemOpen
  \bibfield  {author} {\bibinfo {author} {\bibfnamefont {C.}~\bibnamefont
  {Feng}}, \bibinfo {author} {\bibfnamefont {A.~J.}\ \bibnamefont {Schultz}},
  \bibinfo {author} {\bibfnamefont {V.}~\bibnamefont {Chaudhary}}, \ and\
  \bibinfo {author} {\bibfnamefont {D.~A.}\ \bibnamefont {Kofke}},\ }\href
  {\doibase 10.1063/1.4927339} {\bibfield  {journal} {\bibinfo  {journal} {The
  Journal of Chemical Physics}\ }\textbf {\bibinfo {volume} {143}},\ \bibinfo
  {eid} {044504} (\bibinfo {year} {2015})}\BibitemShut {NoStop}%
\bibitem [{\citenamefont {Schultz}\ and\ \citenamefont
  {Kofke}(2009{\natexlab{b}})}]{Schultz_2009}%
  \BibitemOpen
  \bibfield  {author} {\bibinfo {author} {\bibfnamefont {A.~J.}\ \bibnamefont
  {Schultz}}\ and\ \bibinfo {author} {\bibfnamefont {D.~A.}\ \bibnamefont
  {Kofke}},\ }\href {\doibase 10.1063/1.3148379} {\bibfield  {journal}
  {\bibinfo  {journal} {Journal of Chemical Physics}\ }\textbf {\bibinfo
  {volume} {130}},\ \bibinfo {pages} {224104} (\bibinfo {year}
  {2009}{\natexlab{b}})}\BibitemShut {NoStop}%
\bibitem [{\citenamefont {Vargas}\ \emph {et~al.}(2001)\citenamefont {Vargas},
  \citenamefont {Mu\~noz},\ and\ \citenamefont {Rodriguez}}]{Vargas_2001}%
  \BibitemOpen
  \bibfield  {author} {\bibinfo {author} {\bibfnamefont {P.}~\bibnamefont
  {Vargas}}, \bibinfo {author} {\bibfnamefont {E.}~\bibnamefont {Mu\~noz}}, \
  and\ \bibinfo {author} {\bibfnamefont {L.}~\bibnamefont {Rodriguez}},\ }\href
  {\doibase 10.1016/S0378-4371(00)00362-9} {\bibfield  {journal} {\bibinfo
  {journal} {Physica A: Statistical Mechanics and its Applications}\ }\textbf
  {\bibinfo {volume} {290}},\ \bibinfo {pages} {92} (\bibinfo {year}
  {2001})}\BibitemShut {NoStop}%
\bibitem [{\citenamefont {Eu}(2009)}]{Eu_2009}%
  \BibitemOpen
  \bibfield  {author} {\bibinfo {author} {\bibfnamefont {B.~C.}\ \bibnamefont
  {Eu}},\ }\href@noop {} {\bibfield  {journal} {\bibinfo  {journal} {ArXiv
  e-prints}\ } (\bibinfo {year} {2009})},\ \Eprint
  {http://arxiv.org/abs/0909.3326} {arXiv:0909.3326 [physics.chem-ph]}
  \BibitemShut {NoStop}%
\bibitem [{\citenamefont {Mamedov}\ and\ \citenamefont
  {Somuncu}(2014)}]{Mamedov_2014}%
  \BibitemOpen
  \bibfield  {author} {\bibinfo {author} {\bibfnamefont {B.}~\bibnamefont
  {Mamedov}}\ and\ \bibinfo {author} {\bibfnamefont {E.}~\bibnamefont
  {Somuncu}},\ }\href {\doibase 10.1016/j.molstruc.2014.04.006} {\bibfield
  {journal} {\bibinfo  {journal} {Journal of Molecular Structure}\ }\textbf
  {\bibinfo {volume} {1068}},\ \bibinfo {pages} {164} (\bibinfo {year}
  {2014})}\BibitemShut {NoStop}%
\bibitem [{\citenamefont {Glasser}(2002)}]{Glasser_2002}%
  \BibitemOpen
  \bibfield  {author} {\bibinfo {author} {\bibfnamefont {M.}~\bibnamefont
  {Glasser}},\ }\href {\doibase 10.1016/S0375-9601(02)00814-9} {\bibfield
  {journal} {\bibinfo  {journal} {Physics Letters A}\ }\textbf {\bibinfo
  {volume} {300}},\ \bibinfo {pages} {381} (\bibinfo {year}
  {2002})}\BibitemShut {NoStop}%
\bibitem [{\citenamefont {Gonz\'alez-Calder\'on}\ and\ \citenamefont
  {Rocha-Ichante}(2015)}]{GonzalezCalderon_2015}%
  \BibitemOpen
  \bibfield  {author} {\bibinfo {author} {\bibfnamefont {A.}~\bibnamefont
  {Gonz\'alez-Calder\'on}}\ and\ \bibinfo {author} {\bibfnamefont
  {A.}~\bibnamefont {Rocha-Ichante}},\ }\href {\doibase 10.1063/1.4905663}
  {\bibfield  {journal} {\bibinfo  {journal} {The Journal of Chemical Physics}\
  }\textbf {\bibinfo {volume} {142}},\ \bibinfo {eid} {034305} (\bibinfo {year}
  {2015})}\BibitemShut {NoStop}%
\bibitem [{\citenamefont {Mayer}\ and\ \citenamefont
  {Mayer}(1940)}]{Mayer1940}%
  \BibitemOpen
  \bibfield  {author} {\bibinfo {author} {\bibfnamefont {J.~E.}\ \bibnamefont
  {Mayer}}\ and\ \bibinfo {author} {\bibfnamefont {M.~G.}\ \bibnamefont
  {Mayer}},\ }\href@noop {} {\emph {\bibinfo {title} {Statistical Mechanics}}}\
  (\bibinfo  {publisher} {Wiley},\ \bibinfo {address} {New York},\ \bibinfo
  {year} {1940})\BibitemShut {NoStop}%
\bibitem [{\citenamefont {Hill}(1956)}]{Hill1956}%
  \BibitemOpen
  \bibfield  {author} {\bibinfo {author} {\bibfnamefont {T.~L.}\ \bibnamefont
  {Hill}},\ }\href@noop {} {\emph {\bibinfo {title} {Statistical Mechanics}}}\
  (\bibinfo  {publisher} {Dover},\ \bibinfo {address} {New York},\ \bibinfo
  {year} {1956})\BibitemShut {NoStop}%
\bibitem [{\citenamefont {McQuarrie}(2000)}]{McQuarrie2000}%
  \BibitemOpen
  \bibfield  {author} {\bibinfo {author} {\bibfnamefont {D.~A.}\ \bibnamefont
  {McQuarrie}},\ }\href@noop {} {\emph {\bibinfo {title} {Statistical
  Mechanics}}}\ (\bibinfo  {publisher} {University Science Books},\ \bibinfo
  {address} {Sausalito},\ \bibinfo {year} {2000})\BibitemShut {NoStop}%
\bibitem [{\citenamefont {Hansen}\ and\ \citenamefont
  {McDonald}(2006)}]{Hansen2006}%
  \BibitemOpen
  \bibfield  {author} {\bibinfo {author} {\bibfnamefont {J.-P.}\ \bibnamefont
  {Hansen}}\ and\ \bibinfo {author} {\bibfnamefont {I.~R.}\ \bibnamefont
  {McDonald}},\ }\href@noop {} {\emph {\bibinfo {title} {Theory of simple
  liquids, 3rd Edition}}}\ (\bibinfo  {publisher} {Academic Press},\ \bibinfo
  {address} {Amsterdam},\ \bibinfo {year} {2006})\BibitemShut {NoStop}%
\bibitem [{\citenamefont {Bellemans}(1962{\natexlab{a}})}]{Bellemans_1962}%
  \BibitemOpen
  \bibfield  {author} {\bibinfo {author} {\bibfnamefont {A.}~\bibnamefont
  {Bellemans}},\ }\href {\doibase 10.1016/0031-8914(62)90037-X} {\bibfield
  {journal} {\bibinfo  {journal} {Physica}\ }\textbf {\bibinfo {volume} {28}},\
  \bibinfo {pages} {493} (\bibinfo {year} {1962}{\natexlab{a}})}\BibitemShut
  {NoStop}%
\bibitem [{\citenamefont {Bellemans}(1962{\natexlab{b}})}]{Bellemans_1962_b}%
  \BibitemOpen
  \bibfield  {author} {\bibinfo {author} {\bibfnamefont {A.}~\bibnamefont
  {Bellemans}},\ }\href {\doibase 10.1016/0031-8914(62)90117-9} {\bibfield
  {journal} {\bibinfo  {journal} {Physica}\ }\textbf {\bibinfo {volume} {28}},\
  \bibinfo {pages} {617} (\bibinfo {year} {1962}{\natexlab{b}})}\BibitemShut
  {NoStop}%
\bibitem [{\citenamefont {Bellemans}(1963)}]{Bellemans_1963}%
  \BibitemOpen
  \bibfield  {author} {\bibinfo {author} {\bibfnamefont {A.}~\bibnamefont
  {Bellemans}},\ }\href {\doibase 10.1016/S0031-8914(63)80167-6} {\bibfield
  {journal} {\bibinfo  {journal} {Physica}\ }\textbf {\bibinfo {volume} {29}},\
  \bibinfo {pages} {548} (\bibinfo {year} {1963})}\BibitemShut {NoStop}%
\bibitem [{\citenamefont {Soko{\l}owski}(1977)}]{Sokolowski_1977}%
  \BibitemOpen
  \bibfield  {author} {\bibinfo {author} {\bibfnamefont {S.}~\bibnamefont
  {Soko{\l}owski}},\ }\href {\doibase 10.1007/BF01588931} {\bibfield  {journal}
  {\bibinfo  {journal} {Czechoslovak Journal of Physics}\ }\textbf {\bibinfo
  {volume} {27}},\ \bibinfo {pages} {850} (\bibinfo {year} {1977})}\BibitemShut
  {NoStop}%
\bibitem [{\citenamefont {Soko{\l}owski}\ and\ \citenamefont
  {Stecki}(1979)}]{Sokolowski_1979}%
  \BibitemOpen
  \bibfield  {author} {\bibinfo {author} {\bibfnamefont {S.}~\bibnamefont
  {Soko{\l}owski}}\ and\ \bibinfo {author} {\bibfnamefont {J.}~\bibnamefont
  {Stecki}},\ }\href@noop {} {\bibfield  {journal} {\bibinfo  {journal} {Acta
  Physica Polonica}\ }\textbf {\bibinfo {volume} {55}},\ \bibinfo {pages} {611}
  (\bibinfo {year} {1979})}\BibitemShut {NoStop}%
\bibitem [{\citenamefont {Stecki}\ and\ \citenamefont
  {Soko{\l}owski}(1980)}]{Stecki_1980}%
  \BibitemOpen
  \bibfield  {author} {\bibinfo {author} {\bibfnamefont {J.}~\bibnamefont
  {Stecki}}\ and\ \bibinfo {author} {\bibfnamefont {S.}~\bibnamefont
  {Soko{\l}owski}},\ }\href {\doibase 10.1080/00268978000100291} {\bibfield
  {journal} {\bibinfo  {journal} {Molecular Physics}\ }\textbf {\bibinfo
  {volume} {39}},\ \bibinfo {pages} {343} (\bibinfo {year} {1980})}\BibitemShut
  {NoStop}%
\bibitem [{\citenamefont {Rowlinson}(1986)}]{Rowlinson_1986}%
  \BibitemOpen
  \bibfield  {author} {\bibinfo {author} {\bibfnamefont {J.~S.}\ \bibnamefont
  {Rowlinson}},\ }\href {\doibase 10.1039/F29868201801} {\bibfield  {journal}
  {\bibinfo  {journal} {J. Chem. Soc., Faraday Trans. 2}\ }\textbf {\bibinfo
  {volume} {82}},\ \bibinfo {pages} {1801} (\bibinfo {year}
  {1986})}\BibitemShut {NoStop}%
\bibitem [{\citenamefont {Rowlinson}(1985)}]{Rowlinson_1985}%
  \BibitemOpen
  \bibfield  {author} {\bibinfo {author} {\bibfnamefont {J.~S.}\ \bibnamefont
  {Rowlinson}},\ }\href {\doibase 10.1098/rspa.1985.0108} {\bibfield  {journal}
  {\bibinfo  {journal} {Proceedings of the Royal Society of London. A.
  Mathematical and Physical Sciences}\ }\textbf {\bibinfo {volume} {402}},\
  \bibinfo {pages} {67} (\bibinfo {year} {1985})}\BibitemShut {NoStop}%
\bibitem [{\citenamefont {Evans}\ \emph {et~al.}(2003)\citenamefont {Evans},
  \citenamefont {Roth},\ and\ \citenamefont {Bryk}}]{Evans_2003}%
  \BibitemOpen
  \bibfield  {author} {\bibinfo {author} {\bibfnamefont {R.}~\bibnamefont
  {Evans}}, \bibinfo {author} {\bibfnamefont {R.}~\bibnamefont {Roth}}, \ and\
  \bibinfo {author} {\bibfnamefont {P.}~\bibnamefont {Bryk}},\ }\href
  {http://stacks.iop.org/0295-5075/62/815} {\bibfield  {journal} {\bibinfo
  {journal} {EPL (Europhysics Letters)}\ }\textbf {\bibinfo {volume} {62}},\
  \bibinfo {pages} {815} (\bibinfo {year} {2003})}\BibitemShut {NoStop}%
\bibitem [{\citenamefont {Evans}\ \emph {et~al.}(2004)\citenamefont {Evans},
  \citenamefont {Henderson},\ and\ \citenamefont {Roth}}]{Evans_2004}%
  \BibitemOpen
  \bibfield  {author} {\bibinfo {author} {\bibfnamefont {R.}~\bibnamefont
  {Evans}}, \bibinfo {author} {\bibfnamefont {J.~R.}\ \bibnamefont
  {Henderson}}, \ and\ \bibinfo {author} {\bibfnamefont {R.}~\bibnamefont
  {Roth}},\ }\href {\doibase 10.1063/1.1819316} {\bibfield  {journal} {\bibinfo
   {journal} {The Journal of Chemical Physics}\ }\textbf {\bibinfo {volume}
  {121}},\ \bibinfo {pages} {12074} (\bibinfo {year} {2004})}\BibitemShut
  {NoStop}%
\bibitem [{\citenamefont {Stewart}\ and\ \citenamefont
  {Evans}(2005{\natexlab{a}})}]{Stewart_2005}%
  \BibitemOpen
  \bibfield  {author} {\bibinfo {author} {\bibfnamefont {M.~C.}\ \bibnamefont
  {Stewart}}\ and\ \bibinfo {author} {\bibfnamefont {R.}~\bibnamefont
  {Evans}},\ }\href {http://stacks.iop.org/0953-8984/17/S3499} {\bibfield
  {journal} {\bibinfo  {journal} {Journal of Physics: Condensed Matter}\
  }\textbf {\bibinfo {volume} {17}},\ \bibinfo {pages} {3499} (\bibinfo {year}
  {2005}{\natexlab{a}})}\BibitemShut {NoStop}%
\bibitem [{\citenamefont {Stewart}\ and\ \citenamefont
  {Evans}(2005{\natexlab{b}})}]{Stewart_2005_b}%
  \BibitemOpen
  \bibfield  {author} {\bibinfo {author} {\bibfnamefont {M.~C.}\ \bibnamefont
  {Stewart}}\ and\ \bibinfo {author} {\bibfnamefont {R.}~\bibnamefont
  {Evans}},\ }\href {\doibase 10.1103/PhysRevE.71.011602} {\bibfield  {journal}
  {\bibinfo  {journal} {Phys. Rev. E}\ }\textbf {\bibinfo {volume} {71}},\
  \bibinfo {pages} {011602} (\bibinfo {year} {2005}{\natexlab{b}})}\BibitemShut
  {NoStop}%
\bibitem [{\citenamefont {Reindl}\ \emph {et~al.}(2015)\citenamefont {Reindl},
  \citenamefont {Bier},\ and\ \citenamefont {Dietrich}}]{Reindl_2015}%
  \BibitemOpen
  \bibfield  {author} {\bibinfo {author} {\bibfnamefont {A.}~\bibnamefont
  {Reindl}}, \bibinfo {author} {\bibfnamefont {M.}~\bibnamefont {Bier}}, \ and\
  \bibinfo {author} {\bibfnamefont {S.}~\bibnamefont {Dietrich}},\ }\href
  {\doibase 10.1103/PhysRevE.91.022406} {\bibfield  {journal} {\bibinfo
  {journal} {Phys. Rev. E}\ }\textbf {\bibinfo {volume} {91}},\ \bibinfo
  {pages} {022406} (\bibinfo {year} {2015})}\BibitemShut {NoStop}%
\bibitem [{\citenamefont {Yang}\ \emph {et~al.}(2013)\citenamefont {Yang},
  \citenamefont {Schultz}, \citenamefont {Errington},\ and\ \citenamefont
  {Kofke}}]{Yang_2013}%
  \BibitemOpen
  \bibfield  {author} {\bibinfo {author} {\bibfnamefont {J.~H.}\ \bibnamefont
  {Yang}}, \bibinfo {author} {\bibfnamefont {A.~J.}\ \bibnamefont {Schultz}},
  \bibinfo {author} {\bibfnamefont {J.~R.}\ \bibnamefont {Errington}}, \ and\
  \bibinfo {author} {\bibfnamefont {D.~A.}\ \bibnamefont {Kofke}},\ }\href
  {\doibase 10.1063/1.4798456} {\bibfield  {journal} {\bibinfo  {journal} {The
  Journal of Chemical Physics}\ }\textbf {\bibinfo {volume} {138}},\ \bibinfo
  {eid} {134706} (\bibinfo {year} {2013})}\BibitemShut {NoStop}%
\bibitem [{\citenamefont {Urrutia}(2008)}]{Urrutia_2008}%
  \BibitemOpen
  \bibfield  {author} {\bibinfo {author} {\bibfnamefont {I.}~\bibnamefont
  {Urrutia}},\ }\href {\doibase 10.1007/s10955-008-9513-3} {\bibfield
  {journal} {\bibinfo  {journal} {Journal of Statistical Physics}\ }\textbf
  {\bibinfo {volume} {131}},\ \bibinfo {pages} {597} (\bibinfo {year}
  {2008})}\BibitemShut {NoStop}%
\bibitem [{\citenamefont {Urrutia}(2010)}]{Urrutia_2010b}%
  \BibitemOpen
  \bibfield  {author} {\bibinfo {author} {\bibfnamefont {I.}~\bibnamefont
  {Urrutia}},\ }\href {\doibase 10.1063/1.3469773} {\bibfield  {journal}
  {\bibinfo  {journal} {The Journal of Chemical Physics}\ }\textbf {\bibinfo
  {volume} {133}},\ \bibinfo {eid} {104503} (\bibinfo {year}
  {2010})}\BibitemShut {NoStop}%
\bibitem [{\citenamefont {Allen}\ and\ \citenamefont
  {Tildesley}(1987)}]{Allen1987}%
  \BibitemOpen
  \bibfield  {author} {\bibinfo {author} {\bibfnamefont {M.}~\bibnamefont
  {Allen}}\ and\ \bibinfo {author} {\bibfnamefont {D.~J.}\ \bibnamefont
  {Tildesley}},\ }\href@noop {} {\emph {\bibinfo {title} {Computer Simulation
  of Liquids}}}\ (\bibinfo  {publisher} {Clarendon Press},\ \bibinfo {address}
  {Oxford},\ \bibinfo {year} {1987})\BibitemShut {NoStop}%
\bibitem [{\citenamefont {Shaul}\ \emph {et~al.}(2010)\citenamefont {Shaul},
  \citenamefont {Schultz},\ and\ \citenamefont {Kofke}}]{Shaul_2010}%
  \BibitemOpen
  \bibfield  {author} {\bibinfo {author} {\bibfnamefont {K.~R.~S.}\
  \bibnamefont {Shaul}}, \bibinfo {author} {\bibfnamefont {A.~J.}\ \bibnamefont
  {Schultz}}, \ and\ \bibinfo {author} {\bibfnamefont {D.~A.}\ \bibnamefont
  {Kofke}},\ }\href {\doibase 10.1135/cccc2009113} {\bibfield  {journal}
  {\bibinfo  {journal} {Collect. Czech. Chem. Commun.}\ }\textbf {\bibinfo
  {volume} {75}},\ \bibinfo {pages} {447} (\bibinfo {year} {2010})}\BibitemShut
  {NoStop}%
\bibitem [{\citenamefont {McQuarrie}\ and\ \citenamefont
  {Rowlinson}(1987)}]{McQuarrie_1987}%
  \BibitemOpen
  \bibfield  {author} {\bibinfo {author} {\bibfnamefont {D.~A.}\ \bibnamefont
  {McQuarrie}}\ and\ \bibinfo {author} {\bibfnamefont {J.~S.}\ \bibnamefont
  {Rowlinson}},\ }\href {\doibase 10.1080/00268978700100651} {\bibfield
  {journal} {\bibinfo  {journal} {Molecular Physics}\ }\textbf {\bibinfo
  {volume} {60}},\ \bibinfo {pages} {977} (\bibinfo {year} {1987})}\BibitemShut
  {NoStop}%
\bibitem [{\citenamefont {Urrutia}\ and\ \citenamefont
  {Castelletti}(2011)}]{Urrutia_2011}%
  \BibitemOpen
  \bibfield  {author} {\bibinfo {author} {\bibfnamefont {I.}~\bibnamefont
  {Urrutia}}\ and\ \bibinfo {author} {\bibfnamefont {G.}~\bibnamefont
  {Castelletti}},\ }\href {\doibase 10.1063/1.3544681} {\bibfield  {journal}
  {\bibinfo  {journal} {The Journal of Chemical Physics}\ }\textbf {\bibinfo
  {volume} {134}},\ \bibinfo {eid} {064508} (\bibinfo {year}
  {2011})}\BibitemShut {NoStop}%
\bibitem [{Note1()}]{Note1}%
  \BibitemOpen
  \bibinfo {note} {We have found a typo in $a_{2}$ and $c_{2}$ taken from Ref.
  \cite {Urrutia_2011}. It was amended here in the results of $a_{2}$ and
  $c_{2}$ for square-well interaction.}\BibitemShut {Stop}%
\bibitem [{\citenamefont {Lamarche}\ and\ \citenamefont
  {Leroy}(1990)}]{Lamarche_1990}%
  \BibitemOpen
  \bibfield  {author} {\bibinfo {author} {\bibfnamefont {F.}~\bibnamefont
  {Lamarche}}\ and\ \bibinfo {author} {\bibfnamefont {C.}~\bibnamefont
  {Leroy}},\ }\href {\doibase 10.1016/0010-4655(90)90184-3} {\bibfield
  {journal} {\bibinfo  {journal} {Computer Physics Communications}\ }\textbf
  {\bibinfo {volume} {59}},\ \bibinfo {pages} {359} (\bibinfo {year}
  {1990})}\BibitemShut {NoStop}%
\bibitem [{\citenamefont {Urrutia}(2014)}]{Urrutia_2014}%
  \BibitemOpen
  \bibfield  {author} {\bibinfo {author} {\bibfnamefont {I.}~\bibnamefont
  {Urrutia}},\ }\href {\doibase 10.1103/PhysRevE.89.032122} {\bibfield
  {journal} {\bibinfo  {journal} {Phys. Rev. E}\ }\textbf {\bibinfo {volume}
  {89}},\ \bibinfo {pages} {032122} (\bibinfo {year} {2014})}\BibitemShut
  {NoStop}%
\bibitem [{\citenamefont {Yang}\ \emph {et~al.}(2015)\citenamefont {Yang},
  \citenamefont {Schultz}, \citenamefont {Errington},\ and\ \citenamefont
  {Kofke}}]{Yang_2015}%
  \BibitemOpen
  \bibfield  {author} {\bibinfo {author} {\bibfnamefont {J.~H.}\ \bibnamefont
  {Yang}}, \bibinfo {author} {\bibfnamefont {A.~J.}\ \bibnamefont {Schultz}},
  \bibinfo {author} {\bibfnamefont {J.~R.}\ \bibnamefont {Errington}}, \ and\
  \bibinfo {author} {\bibfnamefont {D.~A.}\ \bibnamefont {Kofke}},\ }\href
  {\doibase 10.1080/00268976.2014.999840} {\bibfield  {journal} {\bibinfo
  {journal} {Molecular Physics}\ }\textbf {\bibinfo {volume} {113}},\ \bibinfo
  {pages} {1179} (\bibinfo {year} {2015})}\BibitemShut {NoStop}%
\bibitem [{\citenamefont {Vliegenthart}\ and\ \citenamefont
  {Lekkerkerker}(2000)}]{Vliegenthart_2000}%
  \BibitemOpen
  \bibfield  {author} {\bibinfo {author} {\bibfnamefont {G.~A.}\ \bibnamefont
  {Vliegenthart}}\ and\ \bibinfo {author} {\bibfnamefont {H.~N.~W.}\
  \bibnamefont {Lekkerkerker}},\ }\href {\doibase 10.1063/1.481106} {\bibfield
  {journal} {\bibinfo  {journal} {The Journal of Chemical Physics}\ }\textbf
  {\bibinfo {volume} {112}},\ \bibinfo {eid} {5364-5369} (\bibinfo {year}
  {2000})}\BibitemShut {NoStop}%
\bibitem [{\citenamefont {Barker}\ and\ \citenamefont
  {Henderson}(1967)}]{Barker_1967_b}%
  \BibitemOpen
  \bibfield  {author} {\bibinfo {author} {\bibfnamefont {J.~A.}\ \bibnamefont
  {Barker}}\ and\ \bibinfo {author} {\bibfnamefont {D.}~\bibnamefont
  {Henderson}},\ }\href {\doibase 10.1063/1.1701689} {\bibfield  {journal}
  {\bibinfo  {journal} {The Journal of Chemical Physics}\ }\textbf {\bibinfo
  {volume} {47}},\ \bibinfo {pages} {4714} (\bibinfo {year}
  {1967})}\BibitemShut {NoStop}%
\bibitem [{\citenamefont {Henderson}\ and\ \citenamefont
  {Barker}(1970)}]{HendersonD_1970}%
  \BibitemOpen
  \bibfield  {author} {\bibinfo {author} {\bibfnamefont {D.}~\bibnamefont
  {Henderson}}\ and\ \bibinfo {author} {\bibfnamefont {J.~A.}\ \bibnamefont
  {Barker}},\ }\href {\doibase 10.1103/PhysRevA.1.1266} {\bibfield  {journal}
  {\bibinfo  {journal} {Phys. Rev. A}\ }\textbf {\bibinfo {volume} {1}},\
  \bibinfo {pages} {1266} (\bibinfo {year} {1970})}\BibitemShut {NoStop}%
\bibitem [{\citenamefont {Tang}\ and\ \citenamefont {Wu}(2003)}]{Tang_2003}%
  \BibitemOpen
  \bibfield  {author} {\bibinfo {author} {\bibfnamefont {Y.}~\bibnamefont
  {Tang}}\ and\ \bibinfo {author} {\bibfnamefont {J.}~\bibnamefont {Wu}},\
  }\href {\doibase h10.1063/1.1607956} {\bibfield  {journal} {\bibinfo
  {journal} {The Journal of Chemical Physics}\ }\textbf {\bibinfo {volume}
  {119}},\ \bibinfo {pages} {7388} (\bibinfo {year} {2003})}\BibitemShut
  {NoStop}%
\bibitem [{\citenamefont {Orea}\ \emph {et~al.}(2015)\citenamefont {Orea},
  \citenamefont {Romero-Mart\'inez}, \citenamefont {Basurto}, \citenamefont
  {Vargas},\ and\ \citenamefont {Odriozola}}]{Orea_2015}%
  \BibitemOpen
  \bibfield  {author} {\bibinfo {author} {\bibfnamefont {P.}~\bibnamefont
  {Orea}}, \bibinfo {author} {\bibfnamefont {A.}~\bibnamefont
  {Romero-Mart\'inez}}, \bibinfo {author} {\bibfnamefont {E.}~\bibnamefont
  {Basurto}}, \bibinfo {author} {\bibfnamefont {C.~A.}\ \bibnamefont {Vargas}},
  \ and\ \bibinfo {author} {\bibfnamefont {G.}~\bibnamefont {Odriozola}},\
  }\href {\doibase 10.1063/1.4926464} {\bibfield  {journal} {\bibinfo
  {journal} {The Journal of Chemical Physics}\ }\textbf {\bibinfo {volume}
  {143}},\ \bibinfo {eid} {024504} (\bibinfo {year} {2015})}\BibitemShut
  {NoStop}%
\bibitem [{\citenamefont {Ashwin}\ and\ \citenamefont
  {Bowles}(2009)}]{Ashwin_2009}%
  \BibitemOpen
  \bibfield  {author} {\bibinfo {author} {\bibfnamefont {S.~S.}\ \bibnamefont
  {Ashwin}}\ and\ \bibinfo {author} {\bibfnamefont {R.~K.}\ \bibnamefont
  {Bowles}},\ }\href {\doibase 10.1103/PhysRevLett.102.235701} {\bibfield
  {journal} {\bibinfo  {journal} {Phys. Rev. Lett.}\ }\textbf {\bibinfo
  {volume} {102}},\ \bibinfo {pages} {235701} (\bibinfo {year}
  {2009})}\BibitemShut {NoStop}%
\bibitem [{\citenamefont {Khordad}(2012)}]{Khordad_2012}%
  \BibitemOpen
  \bibfield  {author} {\bibinfo {author} {\bibfnamefont {R.}~\bibnamefont
  {Khordad}},\ }\href {http://stacks.iop.org/0253-6102/58/i=5/a=23} {\bibfield
  {journal} {\bibinfo  {journal} {Communications in Theoretical Physics}\
  }\textbf {\bibinfo {volume} {58}},\ \bibinfo {pages} {759} (\bibinfo {year}
  {2012})}\BibitemShut {NoStop}%
\bibitem [{\citenamefont {Urrutia}\ and\ \citenamefont
  {Szybisz}(2010)}]{Urrutia_2010}%
  \BibitemOpen
  \bibfield  {author} {\bibinfo {author} {\bibfnamefont {I.}~\bibnamefont
  {Urrutia}}\ and\ \bibinfo {author} {\bibfnamefont {L.}~\bibnamefont
  {Szybisz}},\ }\href {\doibase 10.1063/1.3319560} {\bibfield  {journal}
  {\bibinfo  {journal} {Journal of Mathematical Physics}\ }\textbf {\bibinfo
  {volume} {51}},\ \bibinfo {pages} {033303} (\bibinfo {year}
  {2010})}\BibitemShut {NoStop}%
\bibitem [{\citenamefont {Vliegenthart}\ \emph {et~al.}(1999)\citenamefont
  {Vliegenthart}, \citenamefont {Lodge},\ and\ \citenamefont
  {Lekkerkerker}}]{Vliegenthart_1999}%
  \BibitemOpen
  \bibfield  {author} {\bibinfo {author} {\bibfnamefont {G.~A.}\ \bibnamefont
  {Vliegenthart}}, \bibinfo {author} {\bibfnamefont {J.~F.~M.}\ \bibnamefont
  {Lodge}}, \ and\ \bibinfo {author} {\bibfnamefont {H.~N.~W.}\ \bibnamefont
  {Lekkerkerker}},\ }\href {\doibase 10.1016/S0378-4371(98)00515-9} {\bibfield
  {journal} {\bibinfo  {journal} {Physica A: Statistical Mechanics and its
  Applications}\ }\textbf {\bibinfo {volume} {263}},\ \bibinfo {pages} {378 }
  (\bibinfo {year} {1999})},\ \bibinfo {note} {proceedings of the 20th
  \{IUPAP\} International Conference on Statistical Physics}\BibitemShut
  {NoStop}%
\bibitem [{\citenamefont {P\'erez-Pellitero}\ \emph {et~al.}(2006)\citenamefont
  {P\'erez-Pellitero}, \citenamefont {Ungerer}, \citenamefont {Orkoulas},\ and\
  \citenamefont {Mackie}}]{PerezPellitero_2006}%
  \BibitemOpen
  \bibfield  {author} {\bibinfo {author} {\bibfnamefont {J.}~\bibnamefont
  {P\'erez-Pellitero}}, \bibinfo {author} {\bibfnamefont {P.}~\bibnamefont
  {Ungerer}}, \bibinfo {author} {\bibfnamefont {G.}~\bibnamefont {Orkoulas}}, \
  and\ \bibinfo {author} {\bibfnamefont {A.~D.}\ \bibnamefont {Mackie}},\
  }\href {\doibase 10.1063/1.2227027} {\bibfield  {journal} {\bibinfo
  {journal} {The Journal of Chemical Physics}\ }\textbf {\bibinfo {volume}
  {125}},\ \bibinfo {eid} {054515} (\bibinfo {year} {2006})}\BibitemShut
  {NoStop}%
\bibitem [{\citenamefont {Urrutia}(2015)}]{Urrutia_2015}%
  \BibitemOpen
  \bibfield  {author} {\bibinfo {author} {\bibfnamefont {I.}~\bibnamefont
  {Urrutia}},\ }\href {\doibase 10.1063/1.4922928} {\bibfield  {journal}
  {\bibinfo  {journal} {The Journal of Chemical Physics}\ }\textbf {\bibinfo
  {volume} {142}},\ \bibinfo {eid} {244902} (\bibinfo {year}
  {2015})}\BibitemShut {NoStop}%
\bibitem [{\citenamefont {Helfrich}(1973)}]{Helfrich_1973}%
  \BibitemOpen
  \bibfield  {author} {\bibinfo {author} {\bibfnamefont {W.}~\bibnamefont
  {Helfrich}},\ }\href@noop {} {\bibfield  {journal} {\bibinfo  {journal}
  {Zeitschrift fï¿œr Naturforschung C}\ }\textbf {\bibinfo {volume} {28}},\
  \bibinfo {pages} {693} (\bibinfo {year} {1973})}\BibitemShut {NoStop}%
\bibitem [{\citenamefont {K\"{o}nig}\ \emph {et~al.}(2005)\citenamefont
  {K\"{o}nig}, \citenamefont {Bryk}, \citenamefont {Mecke},\ and\ \citenamefont
  {Roth}}]{Konig_2005}%
  \BibitemOpen
  \bibfield  {author} {\bibinfo {author} {\bibfnamefont {P.~M.}\ \bibnamefont
  {K\"{o}nig}}, \bibinfo {author} {\bibfnamefont {P.}~\bibnamefont {Bryk}},
  \bibinfo {author} {\bibfnamefont {K.~R.}\ \bibnamefont {Mecke}}, \ and\
  \bibinfo {author} {\bibfnamefont {R.}~\bibnamefont {Roth}},\ }\href {\doibase
  10.1209/epl/i2004-10410-4} {\bibfield  {journal} {\bibinfo  {journal}
  {Europhysics Letters}\ }\textbf {\bibinfo {volume} {69}},\ \bibinfo {pages}
  {832} (\bibinfo {year} {2005})}\BibitemShut {NoStop}%
\bibitem [{\citenamefont {Henderson}(1983)}]{Henderson_1983}%
  \BibitemOpen
  \bibfield  {author} {\bibinfo {author} {\bibfnamefont {J.~R.}\ \bibnamefont
  {Henderson}},\ }\href {\doibase 10.1080/00268978300102661} {\bibfield
  {journal} {\bibinfo  {journal} {Molecular Physics}\ }\textbf {\bibinfo
  {volume} {50}},\ \bibinfo {pages} {741} (\bibinfo {year} {1983})}\BibitemShut
  {NoStop}%
\bibitem [{\citenamefont {Stecki}\ and\ \citenamefont
  {Toxvaerd}(1990)}]{Stecki_1990}%
  \BibitemOpen
  \bibfield  {author} {\bibinfo {author} {\bibfnamefont {J.}~\bibnamefont
  {Stecki}}\ and\ \bibinfo {author} {\bibfnamefont {S.}~\bibnamefont
  {Toxvaerd}},\ }\href {\doibase 10.1063/1.459407} {\bibfield  {journal}
  {\bibinfo  {journal} {The Journal of Chemical Physics}\ }\textbf {\bibinfo
  {volume} {93}},\ \bibinfo {pages} {7342} (\bibinfo {year}
  {1990})}\BibitemShut {NoStop}%
\bibitem [{\citenamefont {Poniewierski}\ and\ \citenamefont
  {Stecki}(1997)}]{Poniewierski_1997}%
  \BibitemOpen
  \bibfield  {author} {\bibinfo {author} {\bibfnamefont {A.}~\bibnamefont
  {Poniewierski}}\ and\ \bibinfo {author} {\bibfnamefont {J.}~\bibnamefont
  {Stecki}},\ }\href {\doibase 10.1063/1.473084} {\bibfield  {journal}
  {\bibinfo  {journal} {The Journal of Chemical Physics}\ }\textbf {\bibinfo
  {volume} {106}},\ \bibinfo {pages} {3358} (\bibinfo {year}
  {1997})}\BibitemShut {NoStop}%
\bibitem [{\citenamefont {Samborski}\ \emph {et~al.}(1993)\citenamefont
  {Samborski}, \citenamefont {Stecki},\ and\ \citenamefont
  {Poniewierski}}]{Samborski_1993}%
  \BibitemOpen
  \bibfield  {author} {\bibinfo {author} {\bibfnamefont {A.}~\bibnamefont
  {Samborski}}, \bibinfo {author} {\bibfnamefont {J.}~\bibnamefont {Stecki}}, \
  and\ \bibinfo {author} {\bibfnamefont {A.}~\bibnamefont {Poniewierski}},\
  }\href {\doibase 10.1063/1.464454} {\bibfield  {journal} {\bibinfo  {journal}
  {The Journal of Chemical Physics}\ }\textbf {\bibinfo {volume} {98}},\
  \bibinfo {pages} {8958} (\bibinfo {year} {1993})}\BibitemShut {NoStop}%
\bibitem [{\citenamefont {Corti}\ \emph {et~al.}(2011)\citenamefont {Corti},
  \citenamefont {Kerr},\ and\ \citenamefont {Torabi}}]{Corti_2011}%
  \BibitemOpen
  \bibfield  {author} {\bibinfo {author} {\bibfnamefont {D.~S.}\ \bibnamefont
  {Corti}}, \bibinfo {author} {\bibfnamefont {K.~J.}\ \bibnamefont {Kerr}}, \
  and\ \bibinfo {author} {\bibfnamefont {K.}~\bibnamefont {Torabi}},\ }\href
  {\doibase 10.1063/1.3609274} {\bibfield  {journal} {\bibinfo  {journal}
  {Journal of Chemical Physics}\ }\textbf {\bibinfo {volume} {135}},\ \bibinfo
  {pages} {024701} (\bibinfo {year} {2011})}\BibitemShut {NoStop}%
\bibitem [{\citenamefont {Hansen-Goos}(2014)}]{HansenGoos_2014}%
  \BibitemOpen
  \bibfield  {author} {\bibinfo {author} {\bibfnamefont {H.}~\bibnamefont
  {Hansen-Goos}},\ }\href {\doibase http://dx.doi.org/10.1063/1.4901110}
  {\bibfield  {journal} {\bibinfo  {journal} {The Journal of Chemical Physics}\
  }\textbf {\bibinfo {volume} {141}},\ \bibinfo {eid} {171101} (\bibinfo {year}
  {2014})}\BibitemShut {NoStop}%
\bibitem [{\citenamefont {Tr\"{o}ster}\ \emph {et~al.}(2012)\citenamefont
  {Tr\"{o}ster}, \citenamefont {Oettel}, \citenamefont {Block}, \citenamefont
  {Virnau},\ and\ \citenamefont {Binder}}]{Troster_2012}%
  \BibitemOpen
  \bibfield  {author} {\bibinfo {author} {\bibfnamefont {A.}~\bibnamefont
  {Tr\"{o}ster}}, \bibinfo {author} {\bibfnamefont {M.}~\bibnamefont {Oettel}},
  \bibinfo {author} {\bibfnamefont {B.~J.}\ \bibnamefont {Block}}, \bibinfo
  {author} {\bibfnamefont {P.}~\bibnamefont {Virnau}}, \ and\ \bibinfo {author}
  {\bibfnamefont {K.}~\bibnamefont {Binder}},\ }\href {\doibase
  10.1063/1.3685221} {\bibfield  {journal} {\bibinfo  {journal} {The Journal of
  Chemical Physics}\ }\textbf {\bibinfo {volume} {136}},\ \bibinfo {eid}
  {064709} (\bibinfo {year} {2012})}\BibitemShut {NoStop}%
\bibitem [{\citenamefont {Baidakov}\ \emph {et~al.}(2009)\citenamefont
  {Baidakov}, \citenamefont {Protsenko},\ and\ \citenamefont
  {Gorbatovskaya}}]{Baidakov_2009}%
  \BibitemOpen
  \bibfield  {author} {\bibinfo {author} {\bibfnamefont {V.~G.}\ \bibnamefont
  {Baidakov}}, \bibinfo {author} {\bibfnamefont {S.~P.}\ \bibnamefont
  {Protsenko}}, \ and\ \bibinfo {author} {\bibfnamefont {G.}~\bibnamefont
  {Gorbatovskaya}},\ }\href {\doibase 10.1134/S1061933X09040012} {\bibfield
  {journal} {\bibinfo  {journal} {Colloid Journal}\ }\textbf {\bibinfo {volume}
  {71}},\ \bibinfo {pages} {437} (\bibinfo {year} {2009})}\BibitemShut
  {NoStop}%
\bibitem [{\citenamefont {Wilhelmsen}\ \emph {et~al.}(2015)\citenamefont
  {Wilhelmsen}, \citenamefont {Bedeaux},\ and\ \citenamefont
  {Reguera}}]{Wilhelmsen_2015}%
  \BibitemOpen
  \bibfield  {author} {\bibinfo {author} {\bibfnamefont {O.}~\bibnamefont
  {Wilhelmsen}}, \bibinfo {author} {\bibfnamefont {D.}~\bibnamefont {Bedeaux}},
  \ and\ \bibinfo {author} {\bibfnamefont {D.}~\bibnamefont {Reguera}},\ }\href
  {\doibase 10.1063/1.4907588} {\bibfield  {journal} {\bibinfo  {journal} {The
  Journal of Chemical Physics}\ }\textbf {\bibinfo {volume} {142}},\ \bibinfo
  {eid} {064706} (\bibinfo {year} {2015})}\BibitemShut {NoStop}%
\bibitem [{\citenamefont {Giessen}\ and\ \citenamefont
  {Blokhuis}(2009)}]{vanGiessen_2009}%
  \BibitemOpen
  \bibfield  {author} {\bibinfo {author} {\bibfnamefont {A.~E.~v.}\
  \bibnamefont {Giessen}}\ and\ \bibinfo {author} {\bibfnamefont {E.~M.}\
  \bibnamefont {Blokhuis}},\ }\href {\doibase 10.1063/1.3253685} {\bibfield
  {journal} {\bibinfo  {journal} {The Journal of Chemical Physics}\ }\textbf
  {\bibinfo {volume} {131}},\ \bibinfo {eid} {164705} (\bibinfo {year}
  {2009})}\BibitemShut {NoStop}%
\bibitem [{\citenamefont {Blokhuis}(2013)}]{Blokhuis_2013}%
  \BibitemOpen
  \bibfield  {author} {\bibinfo {author} {\bibfnamefont {E.~M.}\ \bibnamefont
  {Blokhuis}},\ }\href {\doibase 10.1103/PhysRevE.87.022401} {\bibfield
  {journal} {\bibinfo  {journal} {Phys. Rev. E}\ }\textbf {\bibinfo {volume}
  {87}},\ \bibinfo {pages} {022401} (\bibinfo {year} {2013})}\BibitemShut
  {NoStop}%
\bibitem [{\citenamefont {Schmitz}\ \emph
  {et~al.}(2014{\natexlab{a}})\citenamefont {Schmitz}, \citenamefont {Virnau},\
  and\ \citenamefont {Binder}}]{Schmitz_2014}%
  \BibitemOpen
  \bibfield  {author} {\bibinfo {author} {\bibfnamefont {F.}~\bibnamefont
  {Schmitz}}, \bibinfo {author} {\bibfnamefont {P.}~\bibnamefont {Virnau}}, \
  and\ \bibinfo {author} {\bibfnamefont {K.}~\bibnamefont {Binder}},\ }\href
  {\doibase 10.1103/PhysRevE.90.012128} {\bibfield  {journal} {\bibinfo
  {journal} {Phys. Rev. E}\ }\textbf {\bibinfo {volume} {90}},\ \bibinfo
  {pages} {012128} (\bibinfo {year} {2014}{\natexlab{a}})}\BibitemShut
  {NoStop}%
\bibitem [{\citenamefont {Schmitz}\ \emph
  {et~al.}(2014{\natexlab{b}})\citenamefont {Schmitz}, \citenamefont {Virnau},\
  and\ \citenamefont {Binder}}]{Schmitz_2014b}%
  \BibitemOpen
  \bibfield  {author} {\bibinfo {author} {\bibfnamefont {F.}~\bibnamefont
  {Schmitz}}, \bibinfo {author} {\bibfnamefont {P.}~\bibnamefont {Virnau}}, \
  and\ \bibinfo {author} {\bibfnamefont {K.}~\bibnamefont {Binder}},\ }\href
  {\doibase 10.1103/PhysRevLett.112.125701} {\bibfield  {journal} {\bibinfo
  {journal} {Phys. Rev. Lett.}\ }\textbf {\bibinfo {volume} {112}},\ \bibinfo
  {pages} {125701} (\bibinfo {year} {2014}{\natexlab{b}})}\BibitemShut
  {NoStop}%
\bibitem [{Note2()}]{Note2}%
  \BibitemOpen
  \bibinfo {note} {In Ref.\cite {Lamarche_1990} we found a typo in the
  definition of the elliptic integral $E$. Below Eq. (3) it that work, the
  expression should be $E=\intop \nolimits _{0}^{1}\left (\protect \sqrt
  {1-k^{2}z^{2}}/\protect \sqrt {1-z^{2}}\right )dz$.}\BibitemShut {Stop}%
\bibitem [{\citenamefont {Abramowitz}\ and\ \citenamefont
  {Stegun}(1972)}]{Abramowitz1972}%
  \BibitemOpen
  \bibfield  {author} {\bibinfo {author} {\bibfnamefont {M.}~\bibnamefont
  {Abramowitz}}\ and\ \bibinfo {author} {\bibfnamefont {I.~A.}\ \bibnamefont
  {Stegun}},\ }\href@noop {} {\emph {\bibinfo {title} {Handbook of Mathematical
  Functions}}}\ (\bibinfo  {publisher} {Dover Publications},\ \bibinfo
  {address} {New York},\ \bibinfo {year} {1972})\BibitemShut {NoStop}%
\end{thebibliography}

%

\appendix

\section{Function $S(r,u)$ for cylindrical walls\label{Apsec:CylWalls}}

The surface term $S(r,u)$ was evaluated by taking the derivative
of the volume of intersection between a cylinder and a sphere which
can be expressed in terms of elliptic integrals.\footnote{In Ref.\cite{Lamarche_1990} we found a typo in the definition of
the elliptic integral $E$. Below Eq. (3) it that work, the expression
should be $E=\int_{0}^{1}\left(\sqrt{1-k^{2}z^{2}}/\sqrt{1-z^{2}}\right)dz$.} For the case $0<r<2R$ we obtained
\[
\begin{array}{c}
S(r,u)=4\pi r^{2}\Theta(-u)+\frac{2r}{\sqrt{R(R+u)}}\left[-4R(R+u)E(Q)+\right.\\
\left(4R^{\text{2}}+2Ru-r^{2}\right)K(Q)+r^{2}\left(\frac{2R}{u}+1\right)\varPi\Bigl(1-\frac{r^{2}}{u^{2}},Q\Bigr)\Bigr]\:,
\end{array}
\]
with $Q=\frac{r^{2}-u^{2}}{4R(R+u)}$ and being $\Theta(x)$ the Heaviside
step function which is $\Theta(x)=1$ if $x>0$ and $\Theta(x)=0$
otherwise. Also, $K(Q)$, $E(Q)$ and $\varPi(x,Q)$ are the complete
elliptic integral of the first, second and third kind, respectively
(here $Q$ is the parameter and $x$ the characteristic).\cite{Abramowitz1972}
Note that $S(r,u)$ is a smooth function at $u=0$, because the discontinuity
in $\Theta(-u)$ compensates with a discontinuity in $\varPi\Bigl(1-\frac{r^{2}}{u^{2}},Q\Bigr)$.
We analyzed the behavior of $S(r,u)$ at large $R$ value by taking
its series expansion. For the case $r>2R$ we found
\[
\begin{array}{c}
S(r,u)=4\pi r^{2}\Theta(u)+\frac{4r}{\sqrt{r^{2}-u^{2}}}\biggl\{(u^{2}-r^{2})E\!\left(\frac{1}{Q}\right)+\\
(2R-u)\biggl[u\,K\!\left(\frac{1}{Q}\right)-\frac{r^{2}}{u}\varPi\!\Bigl(\frac{4R(u-R)}{u^{2}},\frac{1}{Q}\Bigr)\biggr]\biggr\}\:.
\end{array}
\]

\section{Functions $I_{\nu,\varepsilon}$ and $C_{q}$\label{Apsec:InuyC} }

Here we analyzed function $I_{\nu,\varepsilon}$ focusing on its
behavior at small $\varepsilon$. For $\nu<1$ $\underset{\varepsilon\rightarrow0}{\lim}I_{\nu,\varepsilon}$
converges to 
\begin{eqnarray}
I_{\nu} & = & \frac{1}{2}z^{\left(\nu+1\right)/2}\left[\Gamma\left(\frac{1-\nu}{2}\right){}_{1}F_{1}\left(\frac{1-\nu}{2},\frac{1}{2},\frac{z}{4}\right)+\right.\nonumber \\
 &  & \left.\sqrt{z}\,\Gamma\left(1-\frac{\nu}{2}\right){}_{1}F_{1}\left(1-\frac{\nu}{2},\frac{3}{2},\frac{z}{4}\right)\right]\:,\label{aeq:Inu}
\end{eqnarray}
where $_{1}F_{1}\left(a,b,x\right)$ is the Kummer's hypergeometric
function. For this case we find the series expansion 
\begin{equation}
I_{\nu,\varepsilon}=I_{\nu}+z\varepsilon^{1-\nu}\!\Bigl(\frac{1}{\nu-1}\!+\!\frac{z\varepsilon}{\nu-2}\!+\!\frac{z\varepsilon^{2}(z-2)}{2(\nu-3)}\!+\!\ldots\Bigr)\label{aeq:Inue}
\end{equation}
Otherwise, if $\nu\geq1$ then $\underset{\varepsilon\rightarrow0}{\lim}I_{\nu,\varepsilon}$
diverges. $I_{1,\varepsilon}$ has been split in a divergent term,
$z\int_{\varepsilon}^{\infty}u^{-1}\exp\left(-zu^{2}\right)du$, and
a non-divergent term, both have been evaluated separately to obtain
\begin{eqnarray}
I_{1,\varepsilon} & = & -\frac{z}{2}\Bigl[\ln\left(z\varepsilon^{2}\right)+\gamma_{e}-{}_{1}F_{1}^{(1,0,0)}\left(0,\frac{1}{2},\frac{z}{4}\right)-\nonumber \\
 &  & \!\!\!\!\pi\text{Erfi}\left(\frac{\sqrt{z}}{2}\right)\Bigr]-\varepsilon z^{2}+\varepsilon^{2}\left(\frac{z^{2}}{2}-\frac{z^{3}}{4}\right)+\ldots\label{aeq:Inu1e}
\end{eqnarray}
Here, $\gamma_{e}$ is the Euler gamma constant, Erfi$(x)$ is the
imaginary error function and $_{1}F_{1}^{(1,0,0)}(0,b,z)$ is $\partial{}_{1}F_{1}(a,b,z)/\partial a$
evaluated at $a=0$ {[}which is equivalent to $_{1}F_{1}^{(1,0,0)}(0,b,x)=b^{-1}x\,{}_{2}F_{2}(1,1;b+1,2;x)${]}.
For $\nu\neq0$ we obtained the recurrence relation
\begin{equation}
\frac{\nu}{z}I_{\nu+1,\varepsilon}=\varepsilon^{-\nu}\exp\left[-z\left(\varepsilon^{2}-\varepsilon\right)\right]+I_{\nu,\varepsilon}-2I_{\nu-1,\varepsilon}\:,\label{aeq:InueRec}
\end{equation}
which combined with Eqs. (\ref{aeq:Inue}) and (\ref{aeq:Inu1e})
enables to obtain the expansion of $I_{\nu,\varepsilon}$ for any
real value $\nu>1,$ that completes our procedure to obtain $I_{\nu,\varepsilon}$
with $\nu\in\mathbb{R}$. For $\nu>1$ Eq. (\ref{aeq:InueRec}) shows
that the divergence is driven by $I_{\nu,\varepsilon}\sim\frac{z\varepsilon^{1-\nu}}{\nu-1}$.

For $0<q<1$ $\underset{\varepsilon\rightarrow0}{\lim}C_{q}(\varepsilon)$
converges to
\begin{eqnarray}
C_{q}(0) & = & \frac{z^{q/2}}{2}\left[\Gamma\left(-\frac{q}{2}\right){}_{1}F_{1}\left(-\frac{q}{2},\frac{1}{2},\frac{z}{4}\right)+\right.\nonumber \\
 &  & \left.\sqrt{z}\,\Gamma\left(\frac{1-q}{2}\right){}_{1}F_{1}\left(\frac{1-q}{2},\frac{3}{2},\frac{z}{4}\right)\right]\:,\label{aeq:Cq0}
\end{eqnarray}
and the series expansion for $C_{q}(\varepsilon)$ is
\begin{equation}
C_{q}(\varepsilon)=C_{q}(0)+z\varepsilon^{1-q}\!\left(\frac{1}{q-1}+\frac{(z-2)\varepsilon}{2(q-2)}+\frac{z(z-6)\varepsilon^{2}}{6(q-3)}+...\right).\label{aeq:Cqe}
\end{equation}
For $q=1$ we separated second term of Eq. (\ref{eq:Cuq}) in several
terms and evaluated each of them, we found
\begin{eqnarray}
C_{1}(\varepsilon) & = & -z\ln(\sqrt{z}\varepsilon)-\sqrt{\pi}\sqrt{z}e^{z/4}\left[1+\text{Erf}\left(\frac{\sqrt{z}}{2}\right)\right]+\nonumber \\
 &  & \frac{z}{2}\!\left[2-\gamma_{e}+\pi\text{Erfi}\!\left(\frac{\sqrt{z}}{2}\right)+{}_{1}F_{1}^{(1,0,0)}\!\!\left(0,\frac{1}{2},\frac{z}{4}\right)\right]\!+\nonumber \\
 &  & \left(1-\frac{z}{2}\right)z\varepsilon+\frac{6-z}{12}z^{2}\varepsilon^{2}+\mathcal{O}\left(\varepsilon^{3}\right)\:,\label{aeq:Cq1e}
\end{eqnarray}
where the divergent term is $-z\ln(\sqrt{z}\varepsilon)$ and Erf$(x)$
is the error function. For $q>1$ we used the simple relation $C_{q}(\varepsilon)=-\frac{1}{q}\varepsilon^{-q}+\frac{1}{z}I_{q+1,\varepsilon}$
(valid for $q>0$) to obtain the following recurrence relation
\begin{eqnarray}
qC_{q}(\varepsilon) & = & \varepsilon^{-q}\left\{ \exp\left[-z\left(\varepsilon^{2}-\varepsilon\right)\right]-1+\frac{z\varepsilon}{q-1}\right\} +\nonumber \\
 &  & zC_{q-1}(\varepsilon)-2I_{q-1,\varepsilon}\:.\label{aeq:CqeRec}
\end{eqnarray}
This relation joined with Eqs. (\ref{aeq:Cqe}, \ref{aeq:Cq1e})
allow to obtain the expansion of $C_{q}(\varepsilon)$ for all $q>0$.
Eq. (\ref{aeq:CqeRec}) shows that the divergence of $C_{q}(\varepsilon)$
for $q>1$ is driven by $C_{q}(\varepsilon)\sim\frac{zq}{q-1}\varepsilon^{1-q}$.
Finally, we found the following interesting property:
\begin{equation}
\underset{q\rightarrow0}{\lim}\:qC_{q}(0)=-1\:,\label{eq:HighTLimit}
\end{equation}
for all $z>0$ (finite values of $T$) that is used in Sec. \ref{sub:Inhom-LJ}
to study the conditions under which LJ systems behaves as HS.

\section{Coefficients for cylindrical walls\label{Apsec:CoeffCyl}}

For cylindrical walls we also analyzed separately the cases $k>6$
and $k=6$. If $k>6$, $x_{2}$ is proportional to $c_{2}$ {[}see
Eqs. (\ref{eq:x2c2}, \ref{eq:c2}){]} and 
\begin{eqnarray*}
d_{2} & = & -\frac{V2\pi}{k}\Delta C_{3/k}+\frac{A\pi}{2k}\Delta C_{4/k}-\frac{L}{R}\frac{\pi^{2}}{32k}\Delta C_{6/k}+\\
 &  & \frac{L}{R^{3}}\frac{\pi^{2}}{1024k}C_{8/k}+\ldots
\end{eqnarray*}
with a series expansion $d_{2}\approx L(2R)^{5-k}$. The coefficient
of this main term in $d_{2}$ was left unevaluated because higher
order functions contribute to this order. The case $k=8$ requires
a special attention because its series gives $d_{2}\approx L\ln R/R^{3}$.

If $k=6$, 
\begin{eqnarray*}
\tilde{x}_{2} & = & -\frac{V\pi}{3}\Delta C_{1/2}+\frac{A\pi}{12}\Delta C_{2/3}+\frac{L}{R}\frac{\pi^{2}}{192}C_{1}(\varepsilon)+\\
 &  & \frac{L}{R^{3}}\frac{\pi^{2}}{1024k}C_{4/3}+\ldots
\end{eqnarray*}
Note that $R^{-3}C_{4/3}(\varepsilon)\approx R^{-1}$ and the same
occurs with higher order terms like $R^{-5}C_{5/3}(\varepsilon)\approx R^{-1}$.
Therefore, we truncated $\tilde{x}_{2}$ to this order which results
in the expression for $x_{2}$ written in Eq. (\ref{eq:x2}) and leaves
$d_{2}$ unevaluated.
\end{document}